\documentclass[a4paper,twocolumn,11pt]{quantumarticle}
\pdfoutput=1
\usepackage[utf8]{inputenc}
\usepackage[english]{babel}
\usepackage[T1]{fontenc}
\usepackage{amsmath}
\usepackage{hyperref}
\usepackage{color,soul}
\usepackage{amssymb}
\usepackage{physics}
\usepackage{xcolor}
\usepackage[numbers,sort&compress]{natbib}

\usepackage{tikz}
\usetikzlibrary{quantikz}
\usepackage{siunitx}

\begin{document}

\definecolor{Pank}{rgb}{1.0,0.1,0.5}
\definecolor{red}{rgb}{1.0,0.0,0.0}
\definecolor{MKPurple}{rgb}{0.45,0.0,0.75}
\newcommand{\lbk}[1]{\color{Pank}[lbk: #1]\normalcolor{}}
\newcommand{\MK}[1]{{\color{MKPurple} #1}}

\newcommand{\addFN}[1]{{\color{blue} #1}}
\newcommand{\FN}[1]{\addFN{\textsf{[FN: #1]}}}

\newtheorem{theorem}{Theorem}
\newtheorem{lemma}{Lemma}

\title{Subsystem self-correction of the GKP qubit.}

\author{Brian Chung Hang Cheung}
\affiliation{Department of Computer Science, University of Copenhagen, Denmark}
\email{chung.hang@di.ku.dk}

\author{Lasse Bjørn Kristensen}
\affiliation{Department of Computer Science, University of Copenhagen, Denmark}

\author{Frederik Nathan}
\affiliation{NNF Quantum Computing Programme, Niels Bohr Institute, University of Copenhagen, Denmark.}

\author{Michael Kastoryano}
\affiliation{Department of Computer Science, University of Copenhagen, Denmark}
\email{mika@di.ku.dk}

\maketitle

\begin{abstract}
  Passive quantum error correction, also known as self-correction, is a holy grail in quantum information science. 
  Recent theoretical advances suggest that the Gottesman-Kitaev-Preskill (GKP) code can exhibit self-correction properties, positioning it as a candidate for the realization of self-correcting quantum memories.
  In this article, we provide a self-contained derivation of the self-correcting behavior of the ideal GKP Hamiltonian, manifested in the Arrhenius type scaling of the logical lifetime, from a quantum-information perspective based on the subsystem code decomposition.
  When coupling through the physical quadrature $q$ and $p$, the detailed-balance jump operators decompose into a dominant part acting only on the gauge subsystem and a boundary term that acts non-trivially on the logical qubit.
  The boundary term is then exponentially suppressed by the Gibbs weights near the edge of the modular cells.
  In contrast to spin-based quantum memories such as the two-dimensional surface code, the GKP Hamiltonian realizes an effective string tension in modular phase space, whereby the energy cost increases as an error approaches the boundary of a logical sector.
\end{abstract}


\section{Introduction}

The protection of quantum information against noise remains one of the central challenges in quantum information science. 
The standard approach is active quantum error correction, in which information is redundantly encoded and repeatedly stabilized by measurements and feedback~\cite{gottesman1997stabilizercodesquantumerror, PhysRevA.52.R2493, Dennis2002}. 
An alternative and conceptually appealing route is \emph{passive} protection, where the physical system itself suppresses logical errors through its native dissipative dynamics~\cite{Alicki2010fourdimention, RevModPhys.88.045005, Brell_2016, PhysRevA.83.042330}. 
This idea is most sharply realized in the theory of self-correcting quantum memories where memory time grows with decreasing temperature, and ideally with a suitable notion of system size, without the need for active intervention. 
While this program has led to profound insights in spin systems~\cite{Brell_2016} and topological phases~\cite{PhysRevA.83.042330}, physically realistic and experimentally accessible self-correcting memories remain elusive.

Continuous variable (CV) quantum information processing provides a different perspective on this problem~\cite{RevModPhys.77.513}. 
Rather than distributing logical information over many two-level systems, bosonic CV codes encode it into one or more infinite-dimensional modes. 
Among these, the Gottesman--Kitaev--Preskill (GKP) code occupies a particularly distinguished position~\cite{gottesman2000encoding}. 
Originally introduced as a grid code in phase space, the GKP encoding combines a rich algebraic structure with unusually strong protection against small displacement errors, and has become a cornerstone of bosonic quantum error correction and fault tolerance. 
Thus, it has inspired a broad experimental and theoretical literature, including
hardware-efficient state-preparation and control schemes~
\cite{Leghtas2013qcMAP,Eickbusch2022FastControl},
autonomous correction protocols~
\cite{Leghtas2013AutonomousMemory,Gertler2021AutonomousQEC},
and architectures for universal fault-tolerant computation~
\cite{Guillaud2019RepetitionCat,Chamberland2022ConcatenatedCat,
Xu2024FaultTolerantBosonic}.
Furthermore, the GKP codes support efficient magic state generation and exponentially robust non-Clifford gates ~\cite{baragiola_all-gaussian_2019,obrien_exponentially_2025,nguyen_fault-tolerant_2025}, addressing one of the central challenges in fault-tolerant quantum computation.
A notable feature of the GKP code is that its ideal code space can be identified with the degenerate low-energy subspace of the stabilizer Hamiltonian~\cite{gottesman2000encoding},
\begin{equation}\label{eqn: GKP Hamiltonian}
H_{\mathrm{GKP}}=-E_J\bigl(\cos(q)+\cos(4\pi p)\bigr),
\end{equation}
where $q$ and $p$ are canonically conjugate dimensionless variables and $E_J$ sets the characteristic energy scale of the stabilizing potential.
The periodic structure in both quadratures, $q$ and $p$, suggests the possibility of intrinsic thermal protection from the energy landscape.

\begin{figure}
    \centering
    \includegraphics[width=1\linewidth]{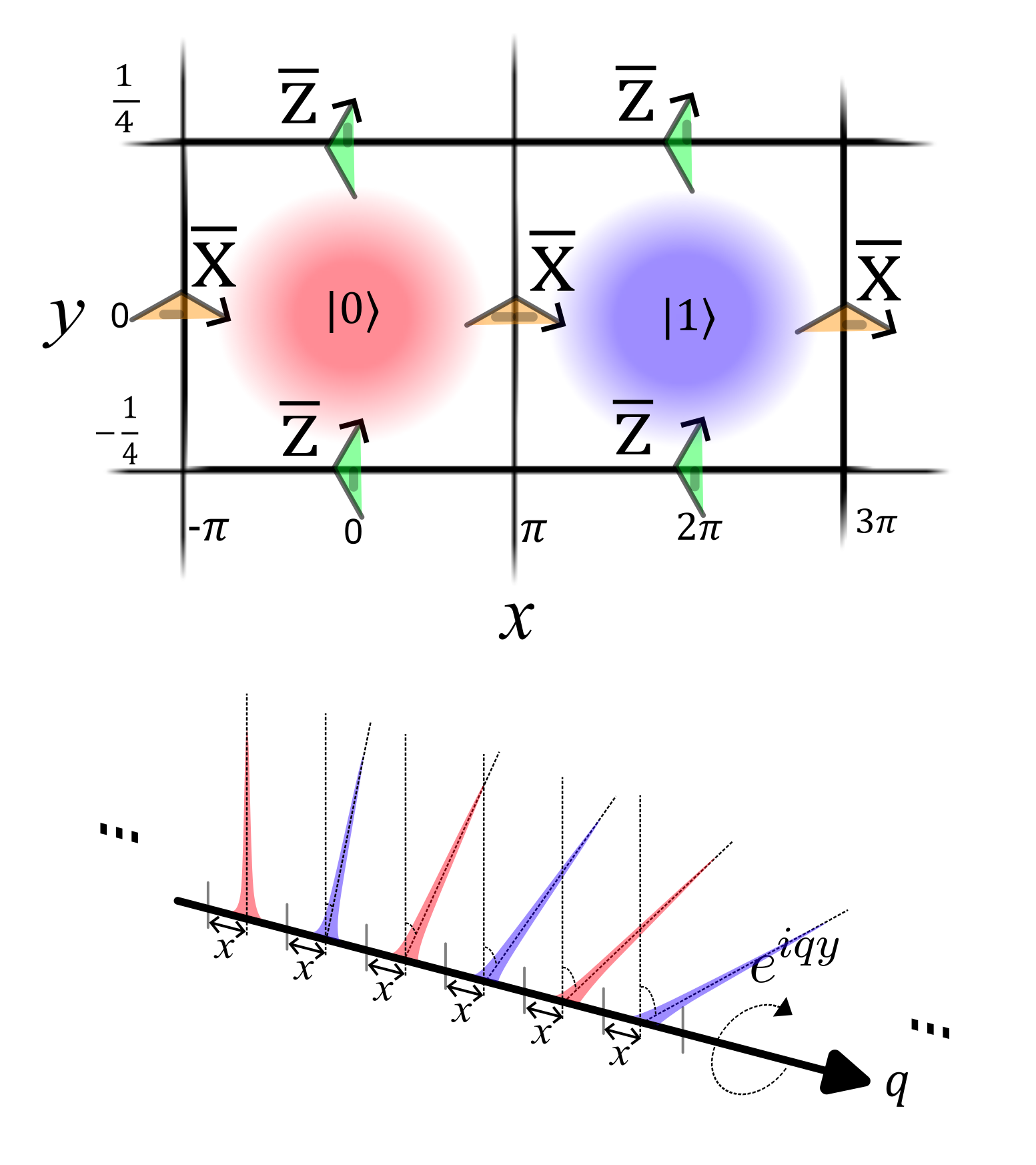}
    \caption{
    Illustration of the subsystem structure of the GKP code in the phase space. States of a single continuous-variable system, described by canonical position and momentum operators $q$ and $p$, can be mapped to those of two quantum variables $x$ and $y$ that are defined on an infinite periodic system, with $x =q\mod 2\pi$ and $y=p\mod 1/2$ representing the (commuting) modular position and momentum of $q$. These modular coordinates parameterise the shift and plane wave amplitude of the Zak basis states.
    The GKP logical state is $|0\rangle$ if $x \in (-\pi, \pi]$  and $y \in (-1/4, 1/4]$, and $|1\rangle$ if $x \in (\pi, 3\pi]$ and $y \in (-1/4, 1/4]$. Horizontal (orange) and vertical (green) boundary crossings correspond to logical $\bar{X}$ and $\bar{Z}$ operations, respectively. $\bar{Z}$ errors correspond to $y$ displacements across the boundaries, and $\bar{X}$ errors to $x$ displacements across the boundaries.}
    \label{fig: two grid diagram}
\end{figure}

Recent work has made this possibility considerably more concrete. 
A particularly important development was the realization that the ideal GKP codeword admits a natural \emph{subsystem} decomposition in the Zak basis, which factors the infinite Hilbert space into a logical subsystem and an auxiliary gauge subsystem~\cite{PRXQuantum.5.010331, PhysRevLett.125.040501, PhysRevA.107.062611}. 
This viewpoint has proven valuable both conceptually and practically: it clarifies the role of modular variables~\cite{PhysRevA.107.062611}, gives a natural operator-algebraic description of logical Pauli operators, and has recently been exploited in proposals for self-correcting GKP implementations and protected GKP gates~\cite{nathan2025selfcorrectinggkpqubitgates,geier2024selfcorrectinggkpqubitsuperconducting}. 
In this representation, the GKP Hamiltonian acts trivially on the logical subsystem and nontrivially only on the gauge degrees of freedom, strongly suggesting that thermalization may primarily affect the gauge sector while leaving the encoded qubit approximately intact.

 To isolate the essential features responsible for this thermal stability, we formulate the dynamics using the exact detailed-balance Lindbladian framework, known as the CKG Lindbladian, introduced in Ref.~\cite{chen2025efficient} and developed further in Refs.~\cite{chen2025efficientexactnoncommutativequantum, Kastoryano2025littlebitofself, gilyen2026quantumgeneralizationsglaubermetropolis, linlin2024ancilla, Lin2025gibbssampler, Lin2025disspipativegroundstate}, which has the exact Gibbs state of the Hamiltonian as a fixed point. 
 While Gibbs states are not exact physical steady states but rather deviate by finite corrections, this model for dissipation is close to physical dissipators (differs only by the addition of a weak counterterm)~\cite{nathan2020universal, nathan2020responsecommentuniversallindblad, nathan2020quantifying}, and thus captures the qualitative aspects of the physics, while being convenient to analyze. 
This choice of CKG Lindbladian is also motivated by two considerations. 
First, detailed-balance Lindbladians provide a mathematically consistent Markovian description of thermalization with the Gibbs state as an exact fixed point. 
Second, they cleanly separate the role of reversibility from the microscopic details of the bath, making them especially well suited for identifying which algebraic properties of the GKP model are responsible for enhanced memory times.

Our analysis shows that the subsystem structure of the ideal GKP code leads to a highly nontrivial Arrhenius-type suppression of logical thermalization. 
When the system couples locally in phase space through the physical quadratures $q$ and $p$, the corresponding detailed-balance jump operators decompose into a dominant part acting only on the gauge subsystem and a boundary term that acts nontrivially on the logical qubit. 
The latter is thermally suppressed by the Gibbs weight of phase-space configurations near the edge of the modular cell, yielding an exponentially small logical dissipation rate. 
For finite-energy states~\cite{gottesman2000encoding} that remain sufficiently close to the ideal GKP code space~\cite{PhysRevLett.125.040501}, the lifetime of the encoded information approximately scales as
\begin{equation}
\tau \sim e^{2\beta E_J},
\end{equation}
up to a prefactor that depends on the bath.
In this sense, the GKP qubit behaves as a thermally stable bosonic memory, with the relevant energy barrier set by the global oscillator energy scale rather than by a local many-body gap. 
We note that the Arrhenius scaling was also derived analytically in Ref.~\cite{nathan2025selfcorrectinggkpqubitgates}. 
\textit{The distinct contribution of the present work is to identify its underlying mechanism from the subsystem code structure of the GKP code.}

This notion of protection is qualitatively different from the one familiar from spin memories. 
For topological stabilizer Hamiltonians such as the surface code~\cite{Dennis2002}, thermal stability is usually discussed in terms of a local excitation gap and an energy barrier associated with growing error strings; in two dimensions, the memory time also has the Arrhenius form~\cite{Chesi_2010, alicki2009}.
For bosonic codes, by contrast, there is no natural thermodynamic limit in which one increases system size while preserving local structure. 
The more fundamental qualitative distinction between the GKP qubit and the surface code is the presence of \textit{string tension}: in the GKP code, the energy cost increases as an error extends toward the boundary of a logical sector, whereas in the surface code, the energy of an error string is determined only by its endpoints and does not grow with its length.

The present paper has two main aims.
The first is pedagogical: to provide a transparent and largely self-contained derivation of the subsystem decomposition of the GKP Hamiltonian in the Zak basis, and to explain how the physical quadratures split into bulk and boundary contributions with distinct logical action. 
The second is conceptual: to show that the thermal robustness of the GKP qubit can be understood as an instance of \emph{subsystem self-correction}, where the environment efficiently thermalizes the gauge sector while only weakly coupling to the encoded logical information. 
We believe this viewpoint helps clarify why the GKP model is special among bosonic encodings and may serve as a useful organizing principle for the design of future passive quantum memories based on subsystem codes~\cite{Chesi_2010}.

\section{Zak basis and the subsystem code decomposition}\label{Section: Zak basis and and the subsystem code decomposition}
The self-correcting properties of the GKP qubit originate in its subsystem code structure~\cite{PRXQuantum.5.010331, PhysRevLett.125.040501, PhysRevA.107.062611}, which allows the full Hilbert space to be factorised into logical and gauge degrees of freedom. In this sense, it can be understood as an instance of \textit{subsystem self-correction}.
To present this decomposition, we first review modular variables~\cite{PhysRevA.107.062611} and the Zak basis~\cite{zak1967zakbasis}, which provide the appropriate representation of the full Hilbert space. 
We then show that the Zak basis is an eigenbasis of the ideal GKP Hamiltonian in Eqn.~\eqref{eqn: GKP Hamiltonian} and that the Hamiltonian acts trivially on the logical subsystem. 
Finally, we derive the subsystem structure of the physical quadratures, $q$ and $p$, which will be used throughout our subsequent analysis of the thermalization of the GKP qubit and its self-correcting properties.
 
\subsection{Modular variables and Zak basis}
\begin{figure}
    \centering
    \includegraphics[width=\linewidth]{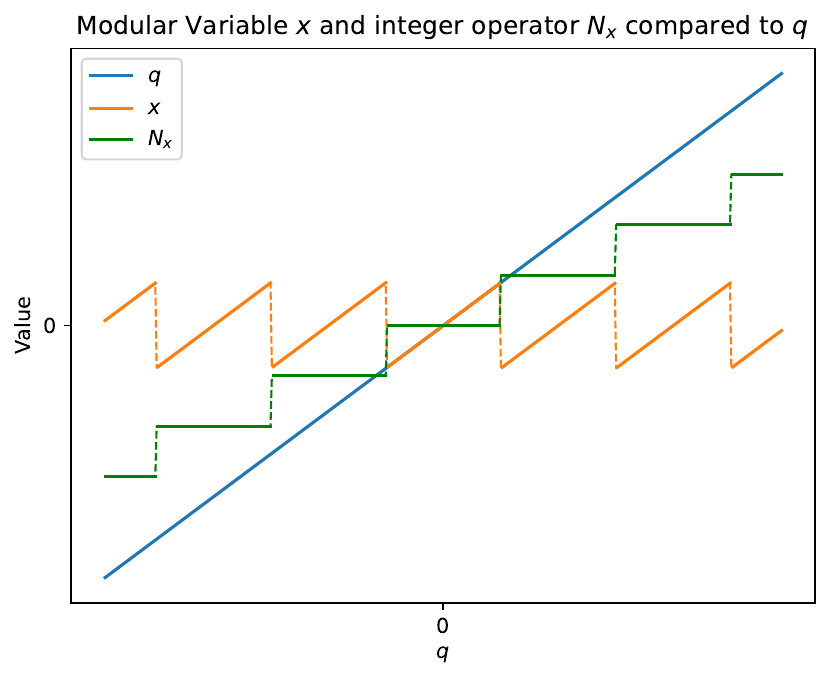}
    \caption{Illustration of the modular variable $x$ and integer operator $N_x$ compared to the position operator $q$. The orange line $x$ takes values within an interval while the green line $N_x$ takes integer values.}
    \label{fig: modular variable}
\end{figure}

We introduce modular variables~\cite{PhysRevA.107.062611} as a natural framework for describing the periodic structure of the GKP code. As illustrated in Fig.~\ref{fig: modular variable}, an operator can be decomposed into a modular component, whose values lie within a fixed interval, and an integer-valued component that labels the corresponding interval.
This decomposition is particularly well suited to the GKP code, whose codewords form a lattice of periodically displaced Dirac peaks, with the logical states $\ket{0}$ and $\ket{1}$ encoded on the even and odd lattice sites, respectively~\cite{gottesman2000encoding}. 
Motivated by this structure, we represent the full Hilbert space of the GKP qubit by partitioning the position and momentum axes into intervals of length $2\pi$ and $1/2$, respectively, and define the associated modular position and momentum operators.
\begin{align}
      q&=x+ 2\pi N_x,\\
    p&=y+\frac{1}{ 2}N_y,
\end{align}
such that modular variables are defined as
\begin{equation}
       x= q\text{ mod }2\pi, \quad y= p\text{ mod }1/2,
\end{equation}
with $x\in(-\pi,\pi]$ and $y\in(-1/4, 1/4]$. Here and throughout, we define the modulo function as
\begin{equation}
    a\text{ mod }b\in(-b/2, b/2].
\end{equation}
Associated with these modular variables are the integer-valued displacement operators
\begin{equation}
N_x = \left\lfloor \frac{q}{2\pi} \right\rceil, \quad N_y = \left\lfloor 2p \right\rceil,
\end{equation}
which label the discrete GKP unit cells that the state has been shifted by.
 As illustrated in Fig.~\ref{fig: two grid diagram}, the parity of $N_x$ determines the logical state along the position axis: even values of $N_x$ correspond to the logical state $\ket{0}$, whereas odd values correspond to $\ket{1}$. Here, $\lfloor a\rceil$ denotes rounding to the nearest integer,
\begin{equation}
    \lfloor a \rceil=a-(a\text{ mod }1).
\end{equation}

The advantage of modular variables lies in the fact that while the position operator and the momentum operator do not commute, i.e. $\comm{q}{p}=i$, their modular counterparts $x$ and $y$ do commute, i.e. $\comm{x}{y}=0$. This commutativity allows the construction of a set of simultaneous eigenbases of $x$ and $y$, also called Zak bases~\cite{zak1967zakbasis, Glancy2006erroranalysis,PRXQuantum.5.010331, PhysRevA.107.062611}.
To make the logical encoding of a qubit explicit, throughout this work we adopt the following convention for the Zak states:
\begin{equation}\label{eqn: zak basis}
    \ket{x,y}=\sqrt{2}\int_{-\infty}^{\infty}dq\, \sum_{n=-\infty}^{\infty}\delta(q-x-4\pi n)e^{ iqy}\ket{q},
\end{equation}
with $|q\rangle$ the eigenstate of the $q$ operator with eigenvalue $q$, 
$(x,y)\in \mathbb{R}^2$, and $n\in \mathbb{Z}$.
It is important to note that the Zak basis is overcomplete and satisfies the following completeness relation:
\begin{align}
\int_{-\pi}^{3\pi} dx \int_{-1/4}^{1/4} dy \ket{x,y}\bra{x, y} = \mathbb{I}.
\end{align}
Furthermore, the Zak basis exhibits quasi-periodicity, characterized by:
\begin{align}
\ket{x+4\pi, y} &= \ket{x, y}, \\
\ket{x, y+1/2} &= e^{ix/2} \ket{x, y},
\end{align}
These properties allow the basis to be interpreted as a torus on the Zak manifold.
As shown by the derivations in Appendix~\ref{Appendix: Zak basis and GKP Hamiltonian}, the Zak bases are eigenstates of the GKP Hamiltonian; this makes them the ideal basis for the analysis of GKP codes.

\subsection{Subsystem code decomposition of the GKP code}
The GKP qubit is encoded in the parity of $N_x$, as illustrated in the upper panel in Fig.~\ref{fig: two grid diagram}: even values of $N_x$ correspond to the logical state $\ket{0}$, whereas odd values correspond to $\ket{1}$. To make this logical degree of freedom explicit, we relabel the Zak basis as~\cite{PRXQuantum.5.010331},
\begin{align}
    \ket{x,y,s}\equiv\ket{x+2\pi s, y},
\end{align}
where $s\in\{0,1\}$ labels the logical states $\ket{0}$ and $\ket{1}$, respectively, while $x\in(-\pi,\pi]$ and $y\in(-1/4,1/4]$. 
Under this new notion, we factorise the full Hilbert space into the logical subsytem $\mathcal{I}$ and the gauge subsystem $\mathcal{S}$~\cite{PRXQuantum.5.010331}, 
\begin{align}
    \mathcal{H}=\mathcal{I}\otimes\mathcal{S}, \quad\ket{x,y, s}&=\ket{s}\otimes\ket{x,y}_G.
\end{align}
Here, the logical component $\ket{s} \in \mathcal{I}$ encodes the logical information, while the gauge component $\ket{x,y}_G \in \mathcal{S}$ specifies the displacement within a GKP unit cell.
Unlike the Zak basis introduced in Eqn.~\eqref{eqn: zak basis}, the gauge Zak basis is defined over the reduced domain $x\in(-\pi,\pi]$ and $y\in(-1/4,1/4]$, and satisfies the completeness relation, 
\begin{align}
\int_{-\pi}^{\pi} dx \int_{-1/4}^{1/4} dy \ket{x,y}_G\bra{x, y}_G = \mathbb{I}_G.
\end{align}
One of the key consequences of the subsystem code decomposition, as shown in Appendix~\ref{Appendix: Zak basis and GKP Hamiltonian}, is that the ideal GKP Hamiltonian in Eqn.~\eqref{eqn: GKP Hamiltonian}, takes the factorized form 
\begin{equation}
    \hat{H}_{\text{GKP}} = \mathbb{I} \otimes \hat{H}_G.  
\end{equation}
More explicitly, the subsystem Zak states satisfy
\begin{equation}
\hat{H}_{\mathrm{GKP}}\ket{x,y,s}
=
E_G(x,y)\ket{x,y,s},
\end{equation}
where the eigenvalue $E_G(x,y)$ is independent of the logical label $s$. The Hamiltonian therefore acts nontrivially only on the gauge subsystem and cannot energetically distinguish between the two logical states.

In contrast to the standard displacement operator approach, the GKP qubit in our setting is encoded through the parity of the integer operators~\cite{PhysRevX.15.011011,nathan2025selfcorrectinggkpqubitgates},
\begin{equation}
    \bar{Z}=(-1)^{N_x}, \quad \bar{X}=(-1)^{N_y}, \quad \bar{Y}=i\bar{X}\bar{Z}.
\end{equation}
These operators satisfy the Pauli algebra and act on the subsystem Zak states according to
\begin{equation}
    \bar{Z}|x,y,s\rangle = (-1)^s|x,y,s\rangle \quad \bar{X}|x,y,s\rangle = |x,y,1-s\rangle. 
\end{equation}
Thus, $\bar Z$ measures the logical label $s$, while $\bar X$ flips it, allowing us to identify the $s$ subsystem as the logical subsystem.
Because these operations do not depend on the gauge coordinates $x$ and $y$, the logical state remains well defined even when the gauge subsystem is in a mixed state~\cite{nathan2025selfcorrectinggkpqubitgates,geier2024selfcorrectinggkpqubitsuperconducting}.

\subsection{Subsystem code decomposition of quadratures}
We begin our analysis by deriving the subsystem code decomposition of the quadrature operators, $q$ and $p$, which provides the necessary foundation for determining the corresponding subsystem structure of the Lindblad jump operators and, subsequently, for analyzing the thermalization of the GKP qubit.
The action of quadratures on the Zak state is 
\begin{align}
    q\ket{x,y}&=(-i\partial_y)\ket{x,y}\label{eq: action of q in zak basis},\\
    p\ket{x,y}&=(y+i\partial_x)\ket{x,y}\label{eq: action of p in zak basis},
\end{align}
as shown in the Appendix.~\ref{Appendix: Action of quadratures on a Zak state}. 
Using integration by parts and the completeness relation on Eqn.~\eqref{eq: action of q in zak basis} and Eqn.~\eqref{eq: action of p in zak basis} yields
\begin{align}
    q&=\int_{-\pi}^{3\pi}dx\,\int_{-1/4}^{1/4}dy\,\ket{x,y}(i\partial_y)\bra{x,y},\\
    p&=\int_{-\pi}^{3\pi}dx\,\int_{-1/4}^{1/4}dy\,\ket{x,y}(y-i\partial_x)\bra{x,y}.
\end{align}
While the $y$ operator can be easily expressed as
\begin{align}
    &\int_{-\pi}^{3\pi}dx\,\int_{-1/4}^{1/4}dy\,\ket{x,y}(y)\bra{x,y}\\
    =&\mathbb{I}\otimes\int_{(-\pi,\pi]}dx\,\int_{-1/4}^{1/4}dy\,y\,\ket{x,y}_G\bra{x,y}_G, 
\end{align}
due to the Zak basis being the eigenstate of modular variables, differential operators require discretization to rewrite in the Zak basis.
Let $x_n=-\pi+n\Delta x$ and $y_m=-\tfrac{1}{4}+m\Delta y$, where $\Delta x=\tfrac{4\pi}{N}$, $\Delta y=\tfrac{1}{2M}$, and $N$ and $M$ are the number of steps, respectively. 
We introduce the discretized Zak basis $\ket{n,y}$ and $\ket{x, m}$
such that 
\begin{align}
    \ket{n,y}&=\sqrt{\Delta x}\ket{x_n,y}\\
    \ket{x,m}&=\sqrt{\Delta y}\ket{x,y_m}.
\end{align}

A notable difference from the continuous Zak basis is the additional normalization factors resulting from the normalization of the Dirac delta to the Kronecker delta
\begin{equation}
    \delta(x_i-x_j)\leftrightarrow\frac{\delta_{i,j}}{\Delta x},
\end{equation}
with a similar relation holding for the $y$-variables. This substitution ensures that the discrete basis remains properly normalized under the discretization of phase space.
In this new basis, the subsystem code decomposition is written as
\begin{align}
    \ket{n,y}&=\ket{0}\otimes\ket{n,y}_G,\label{eqn: discrete subsystem code_0}\\
    \ket{n+N/2,y}&=\ket{1}\otimes\ket{n,y}_G.\label{eqn: discrete subsystem code_1} 
\end{align}
We employ a central finite-difference representation, which preserves Hermiticity, to discretize the operators $-i\partial_x$ and $i\partial_y$, yielding 
\begin{align}
    &\int_{-\pi}^{3\pi}dx\,\int_{-1/4}^{1/4}dy\,\ket{x,y}(-i\partial_x)\bra{x,y}\\
    =&\sum_{n=1}^{N}\Delta x\int_{-1/4}^{1/4}dy\,\frac{\ket{n,y}\bra{n+1,y}-\ket{n,y}\bra{n-1,y}}{2i(\Delta x)^2}\\
    =&\sum_{n=1}^{N}\int_{-1/4}^{1/4}dy\, \frac{\ket{n,y}\bra{n+1,y}-\ket{n,y}\bra{n-1,y}}{2i\Delta x},\label{eq: partial x before subsystem deomposition}\\
    \text{and}\nonumber\\
    &\int_{-\pi}^{3\pi}dx\,\int_{-1/4}^{1/4}dy\,\ket{x,y}(i\partial_y)\bra{x,y}\\
    =&\sum_{m=1}^{M}\Delta y\int_{-\pi}^{3\pi}dx\,\frac{\ket{x,m}\bra{x,m+1}-\ket{x,m}\bra{x,m-1}}{-2i(\Delta y)^2}\\
    =&\sum_{m=1}^{M}\int_{-\pi}^{3\pi}dx\, \frac{\ket{x,m}\bra{x,m+1}-\ket{x,m}\bra{x,m-1}}{-2i\Delta y}.\label{eq: partial y before subsystem deomposition}
\end{align}

We then apply subsystem code decomposition, carefully accounting for terms that cross the boundaries of the GKP unit cell in Figure~\ref{fig: two grid diagram}, as detailed in Appendix~\ref{Appendix: subsystem code decompostion of differetial operator}.
This yields a subsystem code decomposition of the quadratures that reads as, 
\begin{align}
q &=\bar{Z}\otimes\int_{-\pi}^{\pi}dx\,
\frac{e^{-i\frac{x}{2}}\ket{x,M}_G\bra{x,1}_G
}{-2i\Delta y}+h.c.\nonumber\\
&\quad+\mathbb{I}\otimes\sum_{m=1}^{M-1}\int_{-\pi}^{\pi}dx\, 
\frac{\ket{x,m}_G\bra{x,m+1}_G}{-2i\Delta y}+h.c. \label{eq: q in subsystem decomposition}\\
\text{and}\nonumber\\
p &=\bar{X}\otimes\int_{-1/4}^{1/4} dy \,
\frac{\ket{N/2,y}_G\bra{1,y}_G
}{2i \Delta x}+h.c.\nonumber\\
&\quad+\mathbb{I}\otimes\sum_{n=1}^{N/2-1}\int_{-1/4}^{1/4} dy \,
\frac{\ket{n,y}_G\bra{n+1,y}_G}{2i\Delta x}+h.c.\nonumber\\
&\quad+\mathbb{I}\otimes\int_{(-\pi,\pi]}dx\,\int_{-1/4}^{1/4}dy\,
y\,\ket{x,y}_G\bra{x,y}_G\label{eq: p in subsystem decomposition}
\end{align}
These expressions make explicit the logical and gauge contributions of each quadrature operator. 
In particular, they show that logical transitions arise from terms that cross the boundaries between neighboring GKP unit cells. 
This decomposition will provide the basis for identifying the logical action of the Lindblad jump operators and analyzing the thermalization dynamics of the GKP qubit.

\section{On thermal stability of the subsystem GKP code}\label{Sec: Thermal stability}
We will consider the idealized thermalization model given by the CKG Lindbladian \cite{chen2025efficientexactnoncommutativequantum, Kastoryano2025littlebitofself}. While this model is not necessarily the most microscopically accurate system bath coupling model of open quantum systems, it has the benefit of being a mathematically self-consistent Markovian semigroup model of thermalization which pinpoints the role of reversibility (detailed balance) in the thermalization process. Our interest in this work is not to provide the most accurate open system modelling, but rather to identify the key algebraic features of the GKP system which allow it to possess increased thermal stability, as reflected in the Arrhenius scaling with inverse temperature.

The CKG Lindbladian is written as
\begin{equation}\label{eq: CKG lindadian}
    \begin{split}
        \mathcal{L}(\rho) &= -i \left[ Q, \rho \right]\\
    &\quad\quad+ \sum_a L_a \rho L_a^{\dagger} - \frac{1}{2} \left( L_a^{\dagger} L_a \rho + \rho L_a^{\dagger} L_a \right),
    \end{split}
\end{equation}
where
\begin{align}
    L_a = 2 \pi \sum_{k,j} & \sqrt{\gamma(e_k - e_j)} S^a_{k,j} \left| e_k \right> \left< e_j \right|, \label{eq: expression for original jump opertor}\\
    Q = \frac{i}{2} \sum_{i,j} & \tanh \left( \frac{\beta ( e_i - e_j)}{4} \right) \\
    & \times \left< e_i \right| \sum_a L_a^\dagger L_a \left| e_j\right> \left| e_i \right> \left< e_j \right|.
\end{align}
Here, "primal" jump operators $\{S^a\}$ -- $\hat{q}$ and $\hat{p}$ in our case -- are written in terms of the eigenstates $\ket{e_i}$ of the GKP Hamiltonian, with associated eigenvalues $\{e_i\}$,
\begin{align}
    S^a &= \sum_{k,j} S^a_{k,j} \left| e_k \right> \left< e_j \right| \;.
\end{align}
$\gamma$ is the bath autocorrelation function, which needs to satisfy $\gamma(-\omega)=e^{\omega \beta}\gamma(\omega)$. For convenience, we can assume it takes the specific form $\gamma(\omega)=1/(1+e^{\beta \omega})$ of an Ohmic bath. The CKG Linbladian exactly satisfies detailed balance with respect to the Gibbs state $\rho_\beta \propto e^{-\beta H_{GKP}}$; meaning $ \mathcal{L}(\rho_\beta^{1/2}(\cdot)\rho_\beta^{1/2} ) =\rho_\beta^{1/2} \mathcal{L}^\dag(\cdot)\rho_\beta^{1/2}$. In particular, this implies that the Gibbs state is the stationary state of the semigroup $e^{t\mathcal{L}}$.

 We model the survival time as follows. Suppose the logical information can be encoded in a subsystem $\mathcal{I}$, and the rest of the Hilbert space is collected in a "gauge" sector $\mathcal{S}$. In the case of the  GKP qubit, the natural choice is the Zak basis described in Section~\ref{Section: Zak basis and and the subsystem code decomposition}. Then, the logical state $\phi_L$ will be encoded as $\rho_L=\phi_L\otimes \rho_G^\beta$, where $\rho_G^\beta$ is the Gibbs state of the gauge subsystem. The degradation of the encoded state after a specific time $t$ can be bounded as
\begin{equation}
    ||\rho_L(t)-\rho_L||_1= ||\int_0^t \dot{\rho}_L(s)ds||_1\leq t ||\mathcal{L}(\rho_L)||_1
\end{equation}
Hence, the survival time is governed by the quantity $\|\mathcal{L}(\phi_L\otimes\rho_{G}^{\beta})\|_1$, for any bare logical state $\phi_L$.

\subsection{Bounding $\|\mathcal{L}(\rho_L)\|_1$}
\begin{figure}
    \centering
    \includegraphics[width=\linewidth]{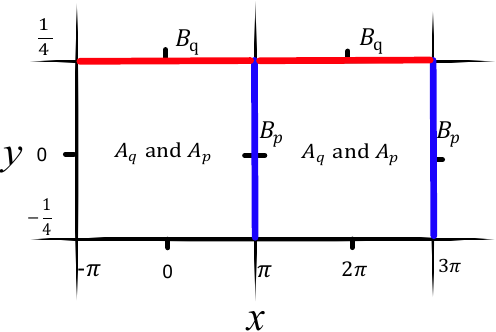}
    \caption{Schematic representation of the subsystem structure of the jump operator $L_p$ and $L_q$, where the operators $A$s act on the bulk of a logical state and the operators $B$s are localized at the boundary between distinct logical states.}
    \label{fig: A and B figure}
\end{figure}

 We show in Appendix~\ref{Appendix: Subsystem code decomposition of jump operators} that the jump operators resulting from the bare $p$ and $q$ bath couplings can be written as
\begin{equation}\label{eqn: jump operator decomposition}
    L_a=\mathbb{I}\otimes A_a+\bar{O}\otimes B_a
\end{equation} 
for $a=q,p$ and $O=Z,X$, respectively. 
For notational simplicity, we henceforth write $A\equiv A_j$ and $B\equiv B_j$ when the corresponding jump operator is clear from context.
In this decomposition, the $A$ operator acts within the bulk of a GKP unit cell, whereas the $B$ operator acts on its boundary, as illustrated in Fig.~\ref{fig: A and B figure}. 
A logical Pauli operator accompanies the latter term and therefore constitutes the source of logical errors induced by thermalization.

We apply the jump-operator decomposition in Eq.~\eqref{eqn: jump operator decomposition} to the CKG Lindbladian in Eq.~\eqref{eq: CKG lindadian}. 
This allows us to separate the Lindbladian into a contribution that acts trivially on the logical subsystem and a remainder that contains all nontrivial logical processes:
\begin{equation}
    \mathcal{L} = \mathbb{I}\otimes\mathcal{A} + \mathcal{R},
\end{equation}
The first contribution, $\mathcal{A}$, acts only on the gauge subsystem and is given by
\begin{equation}
    \mathcal{A}(\cdot) = -i [Q_A,(\cdot)]+A (\cdot) A^\dag -\frac{1}{2} (A^\dag A(\cdot)+(\cdot)A^\dag A).
\end{equation}
The 'correction Hamiltonian' $Q_A$ is given by 
\begin{equation}
   Q_A=\frac{i}{2}\sum_{i,j}\tanh(\frac{\beta(e_i-e_j)}{4})\langle e_i|_G A^\dag A|e_j \rangle_G \ket{e_i}_G\bra{e_j}_G, 
\end{equation}
where $\ket{e}_G$ are the eigenstates of the gauge Hamiltonian $H_G$.
The remainder $\mathcal{R}$ collects all terms involving the boundary contribution $B$ and therefore contains the processes that act nontrivially on the logical subsystem.
Importantly, $\mathcal{R}$ is alone not a Lindbladian. On the other hand, $\mathcal{A}$ is a valid Lindbladian, and even satisfies detailed balance. To see this, note that $\mathcal{A}$ satisfies detailed balance whenever the transition part $A(\cdot)A^\dag$ satisfies detailed balance, as shown in Ref.~\cite{chen2025efficientexactnoncommutativequantum}. But $A(\cdot)A^\dag$ is detailed balanced by virtue of the operator Fourier transform, linearity of $L_j$ and the fact that the GKP Hamiltonian acts trivially on the logical subsystem. Hence, for any logical state $\phi_L$, we get $\mathcal{A}(\rho_G^\beta)=0$.

Hence, bounding the trace norm $||\mathcal{L}(\rho_L)||_1$ reduces to bounding  $||\mathcal{R}(\rho_L)||_1$. In Appendix~\ref{Appendix: proof of lemma 1}, we show that:
 
\begin{lemma}\label{lemma: lemma 1}
    In the setting above, we get
    \begin{equation}
        ||\mathcal{R}(\phi_L\otimes\rho_G^\beta)||_1\leq C ||B \rho_G^\beta||_1+D\|B^\dagger A\rho_G^\beta\|_1,
\end{equation}
where $C$ and $D$ are some constants and $C$ depends on $||A||_{\infty}, ||B||_{\infty}$, which may scale as inverse power of $\Delta x$ and $\Delta y$.  
\end{lemma}
The lemma identifies the trace norm of the boundary terms as the dominant contribution to the logical error. We subsequently show that these boundary contributions exhibit Arrhenius scaling.
More precisely, the Arrhenius factor arises because the boundary operator $B$ probes regions near the boundaries of the GKP unit cell, where the gauge Gibbs state $\rho_G^\beta$ assigns exponentially suppressed weight.

We start with the thermal state of the ideal GKP Hamiltonian:
\begin{equation}
    \rho_{G}^{\beta}=\int_{-\pi}^{\pi} dx\, \int_{-1/4}^{1/4} dy\, \frac{e^{-\epsilon_{x,y}}}{\mathcal{Z}}\dyad{x,y},
\end{equation}
where $\epsilon_{x,y}=-\beta E_J(\cos(x)+\cos(4\pi y))$ and $\mathcal{Z}$ is the partition function.
The thermally activated factor obtained in Appendix~\ref {Appendix: bounding B rho and BArho} for $||B \rho_G^\beta||_1$ and $\|B^\dagger A\rho_G^\beta\|_1$ is
\begin{equation}
    \frac{e^{-\beta E_J}}{2\pi I_0(\beta E_J)},
\end{equation}
and thus
\begin{align}
    \|B^{\dagger}A\rho_G^{\beta}\|_1,\|B\rho_{G}^{\beta}\|_1&\leq 
    K_{\Delta}\frac{e^{-\beta E_J}}{2\pi I_0(\beta E_J)}.
\end{align}
Here, $I_0(\beta E_J)$ is the modified  Bessel function of the first kind and it is known that $I_0(\beta E_J)$ asymptotically approaches $\frac{e^{\beta E_J}}{\sqrt{\beta E_J}}$ as $\beta E_J$ approaches infinity. 
Therefore, we get
\begin{align}
    \|B^{\dagger}A\rho_G^{\beta}\|_1,\|B\rho_{G}^{\beta}\|_1 &\leq K_{\Delta}\sqrt{\beta E_J}e^{-2\beta E_J}.
\end{align}
The prefactor $K_{\Delta}$ depends polynomially on the inverse discretization scales $1/\Delta x$ and $1/\Delta y$.
The divergence of the prefactor does not reflect a physical instability.
Rather, it is an artifact of working with the ideal GKP Hamiltonian, whose eigenstates have infinite energy. 
We adopt this idealized model because its exact subsystem code decomposition provides a transparent framework for isolating the thermodynamic mechanism underlying the Arrhenius scaling of the Lindbladian, which is the central objective of this work.

Strictly speaking, an exact subsystem decomposition is not available for finite-energy GKP states~\cite{PhysRevLett.125.040501}, where the logical and gauge subsystems are no longer perfectly factorized. 
However, in the regime where the finite-energy states closely approximate the ideal code, this decomposition remains approximately valid~\cite{PhysRevLett.125.040501}. 
In this limit, the same physical mechanism persists: boundary processes continue to dominate logical errors, and the resulting dynamics exhibit the same Arrhenius-type suppression. 
This indicates that the \textit{subsystem self-correction} identified here extends to experimentally relevant finite-energy GKP states. A more detailed analysis of the finite-energy case is provided in Appendix~\ref{Appendix: finite energy simulation}.

The divergent prefactor should therefore be understood as a consequence of the ideal-code limit rather than a physically observable effect. In realistic finite-energy implementations, this divergence is regularized.
Consequently, we obtain
\begin{equation}
    \|\mathcal{L}(\phi_L\otimes\rho_{G}^{\beta})\|_1
    \leq K_{\Delta}\sqrt{\beta E_J}e^{-2\beta E_J},
\end{equation}
for both $L_q$ and $L_p$.

\section{Discussion}
In this work, we have identified the mechanism underlying the thermal stability of the GKP qubit. 
The subsystem code decomposition separates the physical Hilbert space into logical and gauge degrees of freedom and reveals that the GKP Hamiltonian acts trivially on the logical subsystem. 
By contrast, the physical quadratures $q$ and $p$, and consequently the corresponding Lindblad jump operators, contain a logically non-trivial gauge contribution that acts on the boundary of the GKP unit cell.  
Logical errors therefore originate from processes that carry the state across the boundary of a GKP unit cell.
{A further analysis starting from a physical dissipation model~\cite{nathan2020universal} is also discussed in Appendix~\ref{Appendix: Corrections from physical dissipation models}.}

Under the specific encoding with gauge Gibbs state, where $\rho_L=\phi_L\otimes\rho^{\beta}_{G}$, exponentially suppressed weights are assigned to these boundary regions. 
As a result, the terms responsible for logical transitions, including quantities of the form $\|B\rho_G^\beta\|_1$ and $\|B^\dagger A\rho_G^\beta\|_1$, exhibit Arrhenius scaling with respect to inverse temperature and the GKP Hamiltonian potential. 
Although this scaling is derived from the physically idealized, infinite-energy GKP Hamiltonian, numerical simulations demonstrate that it remains highly relevant in finite-energy scenarios. Specifically, in the regime where finite-energy states closely approximate the ideal code space, this scaling approximately holds. This yields an approximate Arrhenius-like scaling for the logical lifetime,
\begin{align}
\tau_{\mathrm{GKP}} \sim e^{2\beta E_J}.
\end{align}
The subsystem decomposition thus provides a direct connection between the algebraic structure of the GKP code and its enhanced thermal stability.

Thermal stability is usually discussed in the setting of spin systems~\cite{Alicki2010fourdimention, alicki2009}, where one looks for an energy barrier that grows with the system size; ideally polynomially in the number of degrees of freedom. Such scaling ensures that the survival time of the encoded information grows with the size of the memory. For bosonic encoding, the situation is different, since there is no natural notion of system size that can be increased. 
Instead, the relevant resource is the energy scale of the confining Hamiltonian. In this setting, the best one can hope for is that the survival time grows with the energy scale of the system. The GKP qubit realizes precisely this behavior in which the Arrhenius-like scaling holds.

{\subsection{String tension, thermodynamic transitions, and crossovers}

The GKP stabilizer Hamiltonian realizes a form of string tension in modular phase space. This distinguishes it from conventional two-dimensional topological memories such as the surface code, where the energy of an error string is determined by its endpoints rather than by its length.

Because the GKP and surface-code encodings are geometrically very different, there is no literal correspondence between their respective notions of string length. Nevertheless, a useful analogy can be drawn. In both cases, one may associate an effective distance with an error process and identify a critical distance beyond which recovery fails. In the surface code, this distance is the spatial extent of a string connecting a pair of excitations, or connecting appropriate boundaries in an open geometry. A logical error occurs when such a string becomes topologically nontrivial, for example by winding around the torus. In the single-mode GKP code, the corresponding distance is the accumulated displacement on the Zak torus. A logical error occurs when the displacement crosses a decoding-cell boundary, equivalently when it winds into a neighbouring logical sector.

The existence of a finite energy penalty for creating an error is not, by itself, sufficient for self-correction. What is required is an energy or free-energy barrier that grows with the progress of the logical error. In the present language, this amounts to a nonzero string tension: extending the error string must incur an additional energetic cost. Such a tension generates a restoring force towards the correctable region and can prevent the entropy of long error configurations from overwhelming their energetic suppression at finite temperature.

The two-dimensional surface-code Hamiltonian does not provide such a mechanism. Once a pair of anyonic excitations has been created, the connecting error string can be extended at no further energy cost. Consequently, the maximal energy encountered along a logical-error path remains independent of the system size. More generally, no-go results for local stabilizer Hamiltonians constrain the possibility of obtaining self-correction through a growing string-like energy barrier in spatial dimensions below four~\cite{BravyiTerhal2009,RevModPhys.88.045005}. By contrast, the GKP stabilizer Hamiltonian produces a restoring potential for displacement errors in modular phase space. The energetic cost grows as the displacement moves away from the centre of a decoding cell, thereby realizing an effective string tension.

The comparison is reversed when one considers the critical error length. The surface code admits a thermodynamic limit: by increasing the lattice size, the distance that an error string must traverse to implement a logical operator can be made arbitrarily large. Thus, although the string tension vanishes, the critical string length can diverge with system size. A single-mode GKP code, by contrast, is a zero-dimensional system with no conventional thermodynamic limit. Its decoding cell has finite extent, so the displacement required to cross into a neighbouring logical sector remains finite.

The two encodings therefore exhibit complementary features. The two-dimensional surface code has no string tension but admits a critical error length that grows with system size. The single-mode GKP code has a finite restoring force, or effective string tension, but a finite critical error length. In extended lattice models, the thermodynamic limit permits a sharp distinction between phases with and without self-correction, typically through a finite-temperature transition. For a single GKP mode, no analogous singular thermodynamic transition is available. Instead, one finds a smooth crossover into a regime in which the restoring force strongly suppresses logical errors and the memory lifetime becomes exponentially large in the relevant energetic parameter.

The distinction between a sharp transition and a smooth crossover may, however, be less important operationally than it is conceptually. Any realistic implementation has finite size and is subject to additional imperfections. A lifetime that is finite in principle but exponentially large can therefore provide protection that is, for all practical purposes, indistinguishable from asymptotic self-correction over experimentally relevant timescales.
}
\subsection{Future direction on subsystem self-correction}
More broadly, our results suggest a general route toward thermally stable subsystem codes. The essential ingredients are a Hamiltonian that acts trivially on the logical subsystem, a separation between benign bulk processes and boundary processes that induce logical transitions, and a Gibbs state that exponentially suppresses the relevant boundary region. 
It would be interesting to investigate whether analogous mechanisms arise in other continuous-variable or subsystem encodings, including molecular codes~\cite{albert2020moleule} with GKP-like phase-space structure. 
Related subsystem codes, such as the Bacon--Shor code~\cite{bacon2006baconshor}, may also provide useful settings in which to test the extent to which boundary suppression can protect logical information. 
These directions may help establish more general criteria for self-correction beyond the setting of conventional spin-based quantum memories.

\section*{Acknowledgements}
We thank Giacomo Fregona, Gil Refael, Jonathan Conrad, and Jonas Vinther. We acknowledge support from the Novo Nordisk foundation and from DeiC. FN acknowledges support from  Novo Nordisk Foundation, Grant number NNF22SA0081175, NNF Quantum Computing Programme, from Grant number NNF25OC0102416, and from the the Danish E-infrastructure consortium, grant number 4317-00014B. 

\section*{Author contributions}
All authors contributed to the mathematical development and writing of the manuscript.
Large Language Model were used to assist with proofreading the final paper and the writing part of the code.  

\bibliographystyle{quantum}
\bibliography{bibliography_2}

@article{PRXQuantum.5.010331,
  title = {Stabilizer Subsystem Decompositions for Single- and Multimode {G}ottesman-{K}itaev-{P}reskill Codes},
  author = {Shaw, Mackenzie H. and Doherty, Andrew C. and Grimsmo, Arne L.},
  journal = {PRX Quantum},
  volume = {5},
  issue = {1},
  pages = {010331},
  numpages = {36},
  year = {2024},
  month = {Feb},
  publisher = {American Physical Society},
  doi={https://doi.org/10.1103/PRXQuantum.5.010331}
}

@article{nathan2025selfcorrectinggkpqubitgates,
	title = {Self-{Correcting} {Gottesman}-{Kitaev}-{Preskill} {Qubit} and {Gates} in a {Driven}-{Dissipative} {Circuit}},
	volume = {6},
	issn = {2691-3399},
	doi = {10.1103/ykqb-m52z},
	number = {3},
	journal = {PRX Quantum},
	author = {Nathan, Frederik and O’Brien, Liam and Noh, Kyungjoo and Matheny, Matthew H. and Grimsmo, Arne L. and Jiang, Liang and Refael, Gil},
	month = {sep},
	year = {2025},
	pages = {030352},
}

@misc{geier2024selfcorrectinggkpqubitsuperconducting,
      title={Self-correcting {GKP} qubit in a superconducting circuit with an oscillating voltage bias}, 
      author={Max Geier and Frederik Nathan},
      year={2024},
      eprint={2412.03650},
      archivePrefix={arXiv},
      primaryClass={quant-ph},
      url={https://arxiv.org/abs/2412.03650}, 
}

@article{PhysRevX.15.011011,
  title = {Dissipative Protection of a {GKP} Qubit in a High-Impedance Superconducting Circuit Driven by a Microwave Frequency Comb},
  author = {Sellem, L.-A. and Sarlette, A. and Leghtas, Z. and Mirrahimi, M. and Rouchon, P. and Campagne-Ibarcq, P.},
  journal = {Phys. Rev. X},
  volume = {15},
  issue = {1},
  pages = {011011},
  numpages = {58},
  year = {2025},
  month = {Jan},
  publisher = {American Physical Society},
  doi = {10.1103/PhysRevX.15.011011},
  url = {https://link.aps.org/doi/10.1103/PhysRevX.15.011011}
}

@article{PhysRevLett.125.040501,
  title = {Modular Bosonic Subsystem Codes},
  author = {Pantaleoni, Giacomo and Baragiola, Ben Q. and Menicucci, Nicolas C.},
  journal = {Phys. Rev. Lett.},
  volume = {125},
  issue = {4},
  pages = {040501},
  numpages = {6},
  year = {2020},
  month = {Jul},
  publisher = {American Physical Society},
  doi = {10.1103/PhysRevLett.125.040501},
  url = {https://link.aps.org/doi/10.1103/PhysRevLett.125.040501}
}

@article{PhysRevA.107.062611,
  title = {Zak transform as a framework for quantum computation with the {G}ottesman-{K}itaev-{P}reskill code},
  author = {Pantaleoni, Giacomo and Baragiola, Ben Q. and Menicucci, Nicolas C.},
  journal = {Phys. Rev. A},
  volume = {107},
  issue = {6},
  pages = {062611},
  numpages = {15},
  year = {2023},
  month = {Jun},
  publisher = {American Physical Society},
  doi = {10.1103/PhysRevA.107.062611},
  url = {https://link.aps.org/doi/10.1103/PhysRevA.107.062611}
}

@article{Chesi_2010,
doi = {10.1088/1367-2630/12/2/025013},
url = {https://doi.org/10.1088/1367-2630/12/2/025013},
year = {2010},
month = {feb},
publisher = {},
volume = {12},
number = {2},
pages = {025013},
author = {Chesi, Stefano and Loss, Daniel and Bravyi, Sergey and Terhal, Barbara M},
title = {Thermodynamic stability criteria for a quantum memory based on stabilizer and subsystem codes},
journal = {New Journal of Physics},
}

@article{Kastoryano2025littlebitofself,
  doi = {10.22331/q-2025-08-04-1820},
  url = {https://doi.org/10.22331/q-2025-08-04-1820},
  title = {A little bit of self-correction},
  author = {Kastoryano, Michael J. and Kristensen, Lasse B. and Chen, Chi-Fang and Gily{\'{e}}n, Andras},
  journal = {{Quantum}},
  issn = {2521-327X},
  publisher = {{Verein zur F{\"{o}}rderung des Open Access Publizierens in den Quantenwissenschaften}},
  volume = {9},
  pages = {1820},
  month = aug,
  year = {2025}
}

@article{RevModPhys.88.045005,
  title = {Quantum memories at finite temperature},
  author = {Brown, Benjamin J. and Loss, Daniel and Pachos, Jiannis K. and Self, Chris N. and Wootton, James R.},
  journal = {Rev. Mod. Phys.},
  volume = {88},
  issue = {4},
  pages = {045005},
  numpages = {51},
  year = {2016},
  month = {Nov},
  publisher = {American Physical Society},
  doi = {10.1103/RevModPhys.88.045005},
}

@article{Brell_2016,
doi = {10.1088/1367-2630/18/1/013050},
url = {https://doi.org/10.1088/1367-2630/18/1/013050},
year = {2016},
month = {jan},
publisher = {IOP Publishing},
volume = {18},
number = {1},
pages = {013050},
author = {Brell, Courtney G},
title = {A proposal for self-correcting stabilizer quantum memories in 3 dimensions (or slightly less)},
journal = {New Journal of Physics},
}

@article{PhysRevA.83.042330,
  title = {Local stabilizer codes in three dimensions without string logical operators},
  author = {Haah, Jeongwan},
  journal = {Phys. Rev. A},
  volume = {83},
  issue = {4},
  pages = {042330},
  numpages = {16},
  year = {2011},
  month = {Apr},
  publisher = {American Physical Society},
  doi = {10.1103/PhysRevA.83.042330},
  url = {https://link.aps.org/doi/10.1103/PhysRevA.83.042330}
}

@misc{chen2025efficientexactnoncommutativequantum,
      title={An efficient and exact noncommutative quantum {G}ibbs sampler}, 
      author={Chi-Fang Chen and Michael J. Kastoryano and András Gilyén},
      year={2025},
      eprint={2311.09207},
      archivePrefix={arXiv},
      primaryClass={quant-ph},
      url={https://arxiv.org/abs/2311.09207}, 
}

@article{chen2025efficient,
  title={Efficient quantum thermal simulation},
  author={Chen, Chi-Fang and Kastoryano, Michael and Brand{\~a}o, Fernando GSL and Gily{\'e}n, Andr{\'a}s},
  journal={Nature},
  volume={646},
  number={8085},
  pages={561--566},
  year={2025},
  publisher={Nature Publishing Group UK London},
  doi={https://doi.org/10.1038/s41586-025-09583-x}
}

@misc{obrien_exponentially_2025,
	title = {Exponentially robust non-{Clifford} gate in a driven-dissipative circuit},
	doi = {10.48550/arXiv.2507.19713},
	publisher = {arXiv},
	author = {O'Brien, Liam and Refael, Gil and Nathan, Frederik},
	month = {jul},
	year = {2025},
	note = {arXiv:2507.19713 [quant-ph]},
}

@misc{nguyen_fault-tolerant_2025,
	title = {Fault-{Tolerant} {Non}-{Clifford} {GKP} {Gates} using {Polynomial} {Phase} {Gates} and {On}-{Demand} {Noise} {Biasing}},
	doi = {10.48550/arXiv.2511.20355},
	publisher = {arXiv},
	author = {Nguyen, Minh T. P. and Shaw, Mackenzie H.},
	month = {nov},
	year = {2025},
	note = {arXiv:2511.20355 [quant-ph]},
}

@article{thingna_generalized_2012,
	title = {Generalized {Gibbs} state with modified {Redfield} solution: {Exact} agreement up to second order},
	volume = {136},
	issn = {0021-9606},
	url = {https://doi.org/10.1063/1.4718706},
	doi = {10.1063/1.4718706},
	number = {19},
	journal = {The Journal of Chemical Physics},
	author = {Thingna, Juzar and Wang, Jian-Sheng and Hänggi, Peter},
	month = may,
	year = {2012},
	note = {\_eprint: https://pubs.aip.org/aip/jcp/article-pdf/doi/10.1063/1.4718706/15449334/194110\_1\_online.pdf},
	pages = {194110},
}

@misc{nathan2020responsecommentuniversallindblad,
      title={Response to "Comment on Universal Lindblad Equation for open quantum systems"}, 
      author={Frederik Nathan and Mark S. Rudner},
      year={2020},
      eprint={2011.04574},
      archivePrefix={arXiv},
      primaryClass={quant-ph},
      url={https://arxiv.org/abs/2011.04574}, 
}

@misc{agerskov2026markovianquantummasterequations,
      title={Markovian quantum master equations are exponentially accurate in the weak coupling regime}, 
      author={Johannes Agerskov and Frederik Nathan},
      year={2026},
      eprint={2603.04504},
      archivePrefix={arXiv},
      primaryClass={quant-ph},
      url={https://arxiv.org/abs/2603.04504}, 
}

@article{nathan2020universal,
  title = {Universal Lindblad equation for open quantum systems},
  author = {Nathan, Frederik and Rudner, Mark S.},
  journal = {Phys. Rev. B},
  volume = {102},
  issue = {11},
  pages = {115109},
  numpages = {24},
  year = {2020},
  month = {Sep},
  publisher = {American Physical Society},
  doi = {10.1103/PhysRevB.102.115109},
  url = {https://link.aps.org/doi/10.1103/PhysRevB.102.115109}
}

@article{nathan2020quantifying,
  title = {Quantifying the accuracy of steady states obtained from the universal Lindblad equation},
  author = {Nathan, Frederik and Rudner, Mark S.},
  journal = {Phys. Rev. B},
  volume = {109},
  issue = {20},
  pages = {205140},
  numpages = {9},
  year = {2024},
  month = {May},
  publisher = {American Physical Society},
  doi = {10.1103/PhysRevB.109.205140},
  url = {https://link.aps.org/doi/10.1103/PhysRevB.109.205140}
}

@article{baragiola_all-gaussian_2019,
	title = {All-{Gaussian} {Universality} and {Fault} {Tolerance} with the {Gottesman}-{Kitaev}-{Preskill} {Code}},
	volume = {123},
	doi = {10.1103/PhysRevLett.123.200502},
	number = {20},
	journal = {Physical Review Letters},
	author = {Baragiola, Ben Q. and Pantaleoni, Giacomo and Alexander, Rafael N. and Karanjai, Angela and Menicucci, Nicolas C.},
	year = {2019},
	pages = {200502},
}

@article{alicki2009,
  title={On thermalization in {K}itaev's 2{D} model},
  author={Alicki, Robert and Fannes, Mark and Horodecki, Michal},
  journal={Journal of Physics A: Mathematical and Theoretical},
  volume={42},
  number={6},
  pages={065303},
  year={2009},
  doi={https://doi.org/10.1088/1751-8113/42/6/065303}
}

@article{gottesman2000encoding,
  title={Encoding a qubit in an oscillator},
  author={Gottesman, Daniel and Kitaev, Alexei and Preskill, John},
  journal={arXiv preprint quant-ph/0008040},
  year={2000}
}

@article{Leghtas2013qcMAP,
  title = {Deterministic protocol for mapping a qubit to coherent state superpositions in a cavity},
  author = {Leghtas, Zaki and Kirchmair, Gerhard and Vlastakis, Brian
            and Devoret, Michel H. and Schoelkopf, Robert J.
            and Mirrahimi, Mazyar},
  journal = {Physical Review A},
  volume = {87},
  number = {4},
  pages = {042315},
  year = {2013},
  doi = {10.1103/PhysRevA.87.042315}
}

@article{Eickbusch2022FastControl,
  title = {Fast universal control of an oscillator with weak dispersive coupling to a qubit},
  author = {Eickbusch, Alec and Sivak, Volodymyr and Ding, Andy Z.
            and Elder, Salvatore S. and Jha, Shantanu R.
            and Venkatraman, Jayameenakshi and Royer, Baptiste
            and Girvin, S. M. and Schoelkopf, Robert J.
            and Devoret, Michel H.},
  journal = {Nature Physics},
  volume = {18},
  pages = {1464--1469},
  year = {2022},
  doi = {10.1038/s41567-022-01776-9}
}

@article{Leghtas2013AutonomousMemory,
  title = {Hardware-Efficient Autonomous Quantum Memory Protection},
  author = {Leghtas, Zaki and Kirchmair, Gerhard and Vlastakis, Brian
            and Schoelkopf, Robert J. and Devoret, Michel H.
            and Mirrahimi, Mazyar},
  journal = {Physical Review Letters},
  volume = {111},
  number = {12},
  pages = {120501},
  year = {2013},
  doi = {10.1103/PhysRevLett.111.120501}
}

@article{Gertler2021AutonomousQEC,
  title = {Protecting a bosonic qubit with autonomous quantum error correction},
  author = {Gertler, Jeffrey M. and Baker, Brian and Li, Juliang
            and Shirol, Shruti and Koch, Jens and Wang, Chen},
  journal = {Nature},
  volume = {590},
  pages = {243--248},
  year = {2021},
  doi = {10.1038/s41586-021-03257-0}
}

@article{Guillaud2019RepetitionCat,
  title = {Repetition Cat Qubits for Fault-Tolerant Quantum Computation},
  author = {Guillaud, J{\'e}r{\'e}mie and Mirrahimi, Mazyar},
  journal = {Physical Review X},
  volume = {9},
  number = {4},
  pages = {041053},
  year = {2019},
  doi = {10.1103/PhysRevX.9.041053}
}

@article{Chamberland2022ConcatenatedCat,
  title = {Building a Fault-Tolerant Quantum Computer Using Concatenated Cat Codes},
  author = {Chamberland, Christopher and Noh, Kyungjoo
            and Arrangoiz-Arriola, Patricio and Campbell, Earl T.
            and Hann, Connor T. and Iverson, Joseph and Putterman, Harald
            and Bohdanowicz, Thomas C. and Flammia, Steven T.
            and Keller, Andrew and Refael, Gil and Preskill, John
            and Jiang, Liang and Safavi-Naeini, Amir H.
            and Painter, Oskar and Brand{\~a}o, Fernando G. S. L.},
  journal = {PRX Quantum},
  volume = {3},
  number = {1},
  pages = {010329},
  year = {2022},
  doi = {10.1103/PRXQuantum.3.010329}
}

@article{Xu2024FaultTolerantBosonic,
  title = {Fault-Tolerant Operation of Bosonic Qubits with Discrete-Variable Ancillae},
  author = {Xu, Qian and Zeng, Pei and Xu, Daohong and Jiang, Liang},
  journal = {Physical Review X},
  volume = {14},
  number = {3},
  pages = {031016},
  year = {2024},
  doi = {10.1103/PhysRevX.14.031016}
}

@article{albert2020moleule,
  title = {Robust Encoding of a Qubit in a Molecule},
  author = {Albert, Victor V. and Covey, Jacob P. and Preskill, John},
  journal = {Phys. Rev. X},
  volume = {10},
  issue = {3},
  pages = {031050},
  numpages = {46},
  year = {2020},
  month = {Sep},
  publisher = {American Physical Society},
  doi = {10.1103/PhysRevX.10.031050},
  url = {https://link.aps.org/doi/10.1103/PhysRevX.10.031050}
}

@article{bacon2006baconshor,
  title = {Operator quantum error-correcting subsystems for self-correcting quantum memories},
  author = {Bacon, Dave},
  journal = {Phys. Rev. A},
  volume = {73},
  issue = {1},
  pages = {012340},
  numpages = {13},
  year = {2006},
  month = {Jan},
  publisher = {American Physical Society},
  doi = {10.1103/PhysRevA.73.012340},
  url = {https://link.aps.org/doi/10.1103/PhysRevA.73.012340}
}

@article{zak1967zakbasis,
  title = {Finite Translations in Solid-State Physics},
  author = {Zak, J.},
  journal = {Phys. Rev. Lett.},
  volume = {19},
  issue = {24},
  pages = {1385--1387},
  numpages = {0},
  year = {1967},
  month = {Dec},
  publisher = {American Physical Society},
  doi = {10.1103/PhysRevLett.19.1385},
  url = {https://link.aps.org/doi/10.1103/PhysRevLett.19.1385}
}

@article{Glancy2006erroranalysis,
  title = {Error analysis for encoding a qubit in an oscillator},
  author = {Glancy, S. and Knill, E.},
  journal = {Phys. Rev. A},
  volume = {73},
  issue = {1},
  pages = {012325},
  numpages = {5},
  year = {2006},
  month = {Jan},
  publisher = {American Physical Society},
  doi = {10.1103/PhysRevA.73.012325},
  url = {https://link.aps.org/doi/10.1103/PhysRevA.73.012325}
}

@misc{gilyen2026quantumgeneralizationsglaubermetropolis,
      title={Quantum generalizations of Glauber and Metropolis dynamics}, 
      author={András Gilyén and Chi-Fang Chen and Joao F. Doriguello and Michael J. Kastoryano},
      year={2026},
      eprint={2405.20322},
      archivePrefix={arXiv},
      primaryClass={quant-ph},
      url={https://arxiv.org/abs/2405.20322}, 
}

@article{linlin2024ancilla,
  title = {Single-ancilla ground state preparation via Lindbladians},
  author = {Ding, Zhiyan and Chen, Chi-Fang and Lin, Lin},
  journal = {Phys. Rev. Res.},
  volume = {6},
  issue = {3},
  pages = {033147},
  numpages = {24},
  year = {2024},
  month = {Aug},
  publisher = {American Physical Society},
  doi = {10.1103/PhysRevResearch.6.033147},
  url = {https://link.aps.org/doi/10.1103/PhysRevResearch.6.033147}
}

@article{Lin2025gibbssampler, 
    title={Efficient Quantum gibbs samplers with Kubo–martin–schwinger detailed balance condition}, 
    volume={406}, 
    DOI={10.1007/s00220-025-05235-3}, 
    number={3}, 
    journal={Communications in Mathematical Physics},
    author={Ding, Zhiyan and Li, Bowen and Lin, Lin},
    year={2025}, 
    month={Feb}
}

@article{Lin2025disspipativegroundstate, 
    title={Dissipative ground state preparation in ab initio electronic structure theory}, 
    volume={11}, 
    DOI={10.1038/s41534-025-01124-8}, 
    number={1}, 
    journal={npj Quantum Information}, 
    author={Li, Hao-En and Zhan, Yongtao and Lin, Lin}, 
    year={2025}, 
    month={Nov}
}

@article{BravyiTerhal2009,
  author  = {Bravyi, Sergey and Terhal, Barbara M.},
  title   = {A no-go theorem for a two-dimensional self-correcting
             quantum memory based on stabilizer codes},
  journal = {New Journal of Physics},
  volume  = {11},
  number  = {4},
  pages   = {043029},
  year    = {2009},
  doi     = {10.1088/1367-2630/11/4/043029}
}

@article{Dennis2002,
  author  = {Dennis, Eric and Kitaev, Alexei and Landahl, Andrew
             and Preskill, John},
  title   = {Topological quantum memory},
  journal = {Journal of Mathematical Physics},
  volume  = {43},
  number  = {9},
  pages   = {4452--4505},
  year    = {2002},
  doi     = {10.1063/1.1499754}
}

@article{Alicki2010fourdimention,
author = {Alicki, R. and Horodecki, M. and Horodecki, P. and Horodecki, R.},
title = {On Thermal Stability of Topological Qubit in Kitaev's 4D Model},
journal = {Open Systems \& Information Dynamics},
volume = {17},
number = {01},
pages = {1-20},
year = {2010},
doi = {10.1142/S1230161210000023},
URL = {https://doi.org/10.1142/S1230161210000023},
eprint = {https://doi.org/10.1142/S1230161210000023}

}

@article{RevModPhys.77.513,
  title = {Quantum information with continuous variables},
  author = {Braunstein, Samuel L. and van Loock, Peter},
  journal = {Rev. Mod. Phys.},
  volume = {77},
  issue = {2},
  pages = {513--577},
  numpages = {0},
  year = {2005},
  month = {Jun},
  publisher = {American Physical Society},
  doi = {10.1103/RevModPhys.77.513},
  url = {https://link.aps.org/doi/10.1103/RevModPhys.77.513}
}

@misc{gottesman1997stabilizercodesquantumerror,
      title={Stabilizer Codes and Quantum Error Correction}, 
      author={Daniel Gottesman},
      year={1997},
      eprint={quant-ph/9705052},
      archivePrefix={arXiv},
      primaryClass={quant-ph},
      url={https://arxiv.org/abs/quant-ph/9705052}, 
}

@article{PhysRevA.52.R2493,
  title = {Scheme for reducing decoherence in quantum computer memory},
  author = {Shor, Peter W.},
  journal = {Phys. Rev. A},
  volume = {52},
  issue = {4},
  pages = {R2493(R)--R2496(R)},
  numpages = {0},
  year = {1995},
  month = {Oct},
  publisher = {American Physical Society},
  doi = {10.1103/PhysRevA.52.R2493},
  url = {https://link.aps.org/doi/10.1103/PhysRevA.52.R2493}
}

\onecolumn\newpage
\appendix

\section{Subsystem code decomposition of the GKP Hamiltonian}\label{Appendix: Zak basis and GKP Hamiltonian}

In this Appendix, we formally demonstrate that the Zak basis constitutes the eigenbasis of the GKP Hamiltonian defined in Eqn. \eqref{eqn: GKP Hamiltonian}. Building upon this result, we apply these basis properties to perform a subsystem code decomposition of the GKP Hamiltonian.
Consider the action of $\cos{(q)}$ on the Zak basis, we have
\begin{align}
        \cos{(q)}\ket{x,y}
        &=\sqrt{2}\int_{-\infty}^{\infty}dq\cos{(q)}\sum_{n}\delta(q-x-4\pi n)e^{iqy}\ket{q}.
\end{align}
By expanding the cosine terms, $\cos{(q)}=(e^{iq}+e^{-iq})/2$, and evaluating through the Dirac delta functions, we obtain
\begin{align}
        \cos{(q)}\ket{x,y}&=\frac{\sqrt{2}}{2}\Big(\sum_{n}e^{i(x+4\pi n)y}e^{ix}\ket{x+4\pi n}+\sum_{n}e^{i(x+4\pi n)y}e^{-ix}\ket{x+4\pi n}\Big)\\
        &=\cos{(x)}\ket{x,y}.
\end{align}    
Similarly, let $\ket{p}$ be the eigenstate of the momentum operator and use $\langle p|q\rangle=\frac{1}{\sqrt{2\pi}}e^{-ipq}$, the action of $\cos(4\pi p)\ket{x,y}$ on the Zak basis is expressed as 
\begin{align}
    \cos{(4\pi p)}\ket{x,y}&=\frac{\sqrt{2}}{\sqrt{2\pi}}\int_{-\infty}^{\infty}dq\int_{-\infty}^{\infty}dp\cos{(4\pi p)}\sum_{n}\delta(q-x-4\pi n)e^{iqy}e^{-ipq}\ket{p}.
\end{align}
Again, we expand the cosine terms and use the shifting properties of the Dirac delta function, and we obtain
\begin{align}
    \cos{(4\pi p)}\ket{x,y}&=\frac{\sqrt{2}}{2}\Bigg(\int_{-\infty}^{\infty}dq'\sum_{n'}\delta(q'-x-4\pi n')e^{iq'y}e^{-i4\pi y}\ket{q'}\nonumber\\
        &\quad\quad+\int_{-\infty}^{\infty}dq''\sum_{n''}\delta(q''-x-4\pi n'')e^{iq''y}e^{i4\pi y}\ket{q''}\Bigg)\\
        &=\cos{(4\pi y)}\ket{x,y},
\end{align}
where $q'=q+4\pi$ and $q''=q-4\pi$.

Therefore, the GKP Hamiltonian can be expressed in the diagonal form, 
\begin{equation}
    H_{\text{GKP}}=-E_J\left(\int_{-\pi}^{3\pi}dx\,\int_{-1/4}^{1/4}dy\, (\cos{(x)}+\cos{(4\pi y)})\ket{x,y}\bra{x,y}\right),
\end{equation}
with eigenvalues $-E_J(\cos{(x)}+\cos{(4\pi y)})$.
Applying the subsytem code decomposition, we get
\begin{align}     
    H_{GKP}=&\int_{-\pi}^{3\pi}dx\,\int_{-1/4}^{1/4}dy\, H_{GKP}\ket{x,y}\bra{x,y}\nonumber\\
    =& \int_{-\pi}^{\pi}dx\,\int_{-1/4}^{1/4}dy\, H_{GKP}\ket{x,y}\bra{x,y}+\int_{-\pi}^{\pi}dx\,\int_{-1/4}^{1/4}dy\, H_{GKP}\ket{x+2\pi,y}\bra{x+2\pi,y}\nonumber\\
    =&-E_J(\dyad{0}+\dyad{1})\otimes\int_{-\pi}^{\pi}dx\,\int_{-1/4}^{1/4}dy\, (\cos{(x)}+\cos{(4\pi y)})\ket{x,y}_G\bra{x,y}_G\nonumber\\
    =&\mathbb{I}\otimes H_G,
\end{align}

where $H_G=-E_J\int_{-\pi}^{\pi}dx\,\int_{-1/4}^{1/4}dy\, (\cos{(x)}+\cos{(4\pi y)})\ket{x,y}_G\bra{x,y}_G$. 

\section{Action of quadratures on the Zak state}\label{Appendix: Action of quadratures on a Zak state}

Consider 
\begin{align}
    q\ket{x,y}&=\sqrt{2}\int_{-\infty}^{\infty} dq\, \sum_{n}\delta(q-x-4\pi n)e^{iqy}q\ket{q}\\
    &=\sqrt{2}\int_{-\infty}^{\infty} dq\, \sum_{n}\delta(q-x-4\pi n)(-i\partial_y(e^{iqy}))\ket{q}\\
    &=-i\partial_y\ket{x,y}.
\end{align}
Similarly, 
\begin{align}
    p\ket{x,y}=-\sqrt{2}i\int_{-\infty}^{\infty}dq\,\partial_q\big(\sum_n \delta(q-x-4\pi n)e^{iqy}\big)\ket{q}.
\end{align}
Using the product rule, we get
\begin{align}
    p\ket{x,y}=\sqrt{2}\int_{-\infty}^{\infty}dq \sum_n [y\delta(q-x-4\pi n)-i\partial_q\delta(q-x-4 \pi n)]e^{iqy}\ket{q}.
\end{align}
Using $\partial_xf(x-y)=-\partial_yf(x-y)$, the expression reads as 
\begin{align}
    p\ket{x,y}=\sqrt{2}\int_{-\infty}^{\infty}dq \sum_n [y\delta(q-x-4\pi n)+i\partial_x\delta(q-x-4 \pi n)]e^{iqy}\ket{q}.
\end{align}
We thus have 
\begin{align}
    q\ket{x,y}&=(-i\partial_y)\ket{x,y},\\
    p\ket{x,y}&=(y+i\partial_x)\ket{x,y}.
\end{align}
\section{Subsystem code decomposition of differential operators}\label{Appendix: subsystem code decompostion of differetial operator}
We apply the subsystem code decomposition to differential operators. From Eqn.~\eqref{eq: partial x before subsystem deomposition} and Eqn.~\eqref{eq: partial y before subsystem deomposition}, the $-i\partial_x$ and the $i\partial_y$ operator is written as 
\begin{align}
    \int_{-\pi}^{3\pi}dx\,\int_{-1/4}^{1/4}dy\,\ket{x,y}(-i\partial_x)\bra{x,y}
    &=\sum_{n=1}^{N}\int_{-1/4}^{1/4} dy \, (-i)\frac{\ket{n,y}\bra{n+1,y}-\ket{n,y}\bra{n-1,y}}{2\Delta x}\\
    \int_{-\pi}^{3\pi}dx\,\int_{-1/4}^{1/4}dy\,\ket{x,y}(i\partial_y)\bra{x,y}&=\sum_{m=1}^{M}\int_{-\pi}^{3\pi}dx\, (i)\frac{\ket{x,m}\bra{x,m+1}-\ket{x,m}\bra{x,m-1}}{2\Delta y}
\end{align}
Note that $-\ket{1,y}\bra{0,y}$, $\ket{N,y}\bra{N+1,y}$, $\ket{N/2,y}\bra{N/2+1,y}$ and $-\ket{N/2+1,y}\bra{N/2,y}$ in the $-i\partial_x$ operator cross the boundaries in $x$-direction (see orange arrows in Fig.~\ref{fig: two grid diagram}).
Using the subsystem code decomposition in Eqn.~\eqref{eqn: discrete subsystem code_0}, Eqn.~\eqref{eqn: discrete subsystem code_1} and quasiperiodicity $\ket{x+4\pi,y}=\ket{x,y}$ (i.e. $\ket{0,y}=\ket{N,y}, \ket{1,y}=\ket{N+1,y}$), we get logical jump terms on the boundary, like
\begin{align}
    &\ket{0}\bra{1}\otimes\int_{-1/4}^{1/4} dy \, (-i)\frac{\ket{N/2,y}_G\bra{1,y}_G-\ket{1,y}_G\bra{N/2,y}_G}{2\Delta x},\\
    \text{ and }&\ket{1}\bra{0}\otimes\int_{-1/4}^{1/4} dy \, (-i)\frac{\ket{N/2,y}_G\bra{1,y}_G-\ket{1,y}_G\bra{N/2,y}_G}{2\Delta x}.
\end{align}
In contrast, all other terms together are tensored with $\mathbb{I}$, forming what we define as the bulk operator.
By rearranging terms in the bulk operator, we obtain 
\begin{align}
    &\int_{-\pi}^{3\pi}dx\,\int_{-1/4}^{1/4}dy\,\ket{x,y}(-i\partial_x)\bra{x,y}\nonumber\\
    =&\mathbb{I}\otimes\sum_{n=1}^{N/2-1}\int_{-1/4}^{1/4} dy \, \frac{\ket{n,y}_G\bra{n+1,y}_G-\ket{n+1,y}_G\bra{n,y}_G}{2i\Delta x}\nonumber\\
    &+\bar{X}\otimes\int_{-1/4}^{1/4} dy \, \frac{\ket{N/2,y}_G\bra{1,y}_G-\ket{1,y}_G\bra{N/2,y}_G}{2i\Delta x}.
\end{align}
Similarly, note that $-\ket{x,1}\bra{x,0}$ and $\ket{x,M}\bra{x,M+1}$ in the $i\partial_y$ operator cross the boundary on the $y$-direction (see green arrows in Fig.~\ref{fig: two grid diagram}), resulting in cross-boundary terms like 
\begin{align}
    &\dyad{0}\otimes\int_{-\pi}^{\pi}dx\, (i)\frac{e^{-ix/2}\ket{x,M}_G\bra{x,1}_G-e^{ix/2}\ket{x,1}_G\bra{x,M}_G}{2\Delta y},\\
    \text{ and }-&\dyad{1}\otimes\int_{-\pi}^{\pi}dx\, (i)\frac{e^{-ix/2}\ket{x,M}_G\bra{x,1}_G-e^{ix/2}\ket{x,1}_G\bra{x,M}_G}{2\Delta y}.
\end{align}
where we used the subsystem code decomposition in Eqn.~\eqref{eqn: discrete subsystem code_0}, Eqn.~\eqref{eqn: discrete subsystem code_1} and $\ket{x,y+1/2}=e^{ix/2}\ket{x,y}$ (and thus $\ket{x,0}=e^{-ix/2}\ket{x,M}, \ket{x,M+1}=e^{ix/2}\ket{x,1}$).
Together with the bulk operator and rearranging terms, we get 
\begin{align}
    &\int_{-\pi}^{3\pi}dx\,\int_{-1/4}^{1/4}dy\,\ket{x,y}(-i\partial_y)\bra{x,y}\nonumber\\
    =&\mathbb{I}\otimes\sum_{m=1}^{M-1}\int_{-\pi}^{\pi}dx\, \frac{\ket{x,m}_G\bra{x,m+1}_G-\ket{x,m+1}_G\bra{x,m}_G}{-2i\Delta y}\nonumber\\
    &+\bar{Z}\otimes\int_{-\pi}^{\pi}dx\,\frac{e^{-ix/2}\ket{x,M}_G\bra{x,1}_G-e^{ix/2}\ket{x,1}_G\bra{x,M}_G}{-2i\Delta y}.
\end{align}

\section{Subsystem code decomposition of jump operators}\label{Appendix: Subsystem code decomposition of jump operators}

The subsystem code decomposition of jump operators is obtained by substituting Eqn.~\eqref{eq: q in subsystem decomposition} and Eqn.~\eqref{eq: p in subsystem decomposition} into Eqn.~\eqref{eq: expression for original jump opertor} and Taylor expanding the energy difference, leading to
\begin{align}
    L_q&=\mathbb{I}\otimes A_{q}+\bar{Z}\otimes B_{q},\\
    L_p&=\mathbb{I}\otimes A_{p}+\bar{X}\otimes B_{p},
\end{align}
where 
\begin{align}
    A_q=&\frac{1}{\Delta y}\sum_{m=1}^{M-1}\int_{-\pi}^{\pi}dx\, \left(J_{A,q,m}^{-}\ket{x,m}_G\bra{x,m+1}_G-J_{A,q,m}^{+}\ket{x,m+1}_G\bra{x,m}_G\right),\\
    B_q=&\frac{1}{\Delta y}\int_{-\pi}^{\pi}dx\,\left(J_{B,q}^{+}e^{-\tfrac{ix}{2}}\ket{x,M}_G\bra{x,1}_G-J_{B,q}^{-}e^{\tfrac{ix}{2}}\ket{x,1}_G\bra{x,M}_G\right),\\
    A_p=&\frac{1}{\Delta x}\sum_{n=1}^{N/2-1}\int_{-1/4}^{1/4} dy \, \left(J_{A,p,n}^{-}\ket{n,y}_G\bra{n+1,y}_G-J_{A,p,n}^{+}\ket{n+1,y}_G\bra{n,y}_G\right)\nonumber\\
    &+2\pi\int_{(-\pi,\pi]}dx\,\int_{-1/4}^{1/4}dy\,\sqrt{\gamma(0)}y\,\ket{x,y}_G\bra{x,y}_G,\\
    B_p=&\frac{1}{\Delta x}\int_{-1/4}^{1/4} dy \, \left(J_{B,p}^{+}\ket{N/2,y}_G\bra{1,y}_G-J_{B,p}^{-}\ket{1,y}_G\bra{N/2,y}_G\right),
\end{align}
and the jump rates are expressed as 
\begin{align}
    J_{A,q,m}^{\pm}&=i\pi\sqrt{\gamma(\pm4\pi E_J\Delta y \sin{(4\pi y_m)})},\\
    J_{B,q}^{\pm}&=i\pi \sqrt{\gamma\left(\pm8\pi^2E_J(\Delta y)^2\right)},\\
    J_{A,p,n}^{\pm}&=-i\pi\sqrt{\gamma(\pm E_J\Delta x \sin{(x_n)})},\\
    J_{B,p}^{\pm}&=-i\pi \sqrt{\gamma(\pm E_J(\Delta x)^2/2)}.
\end{align}
Hereafter, we denote $A_p=\tilde{A}_p+D_y$, where we have 
\begin{align}
    \tilde{A}_p=&\frac{1}{\Delta x}\sum_{n=1}^{N/2-1}\int_{-1/4}^{1/4} dy \, \left(J_{A,p,n}^{-}\ket{n,y}_G\bra{n+1,y}_G-J_{A,p,n}^{+}\ket{n+1,y}_G\bra{n,y}_G\right),\\
    D_y=&2\pi\int_{(-\pi,\pi]}dx\,\int_{-1/4}^{1/4}dy\,\sqrt{\gamma(0)}y\,\ket{x,y}_G\bra{x,y}_G.
\end{align}

\section{Proof of Lemma~\ref{lemma: lemma 1} and the bounds}\label{Appendix: proof of lemma 1}
Recall that $\mathcal{A}$ is detailed balanced, and thus $(\mathbb{I} \otimes \mathcal{A})(\phi_L\otimes \rho_G^{\beta})=0$, meaning we have
\begin{equation}
    ||\mathcal{L}(\phi_L\otimes \rho_G^{\beta})||_1 = || \mathcal{R}(\phi_L\otimes \rho_G^{\beta})||_1.
\end{equation}
Given the product structure of the input state and Eqn. (\ref{eqn: jump operator decomposition}), we get 
\begin{align}
    \mathcal{R}(\phi_L\otimes\rho_G^{\beta})=\mathcal{R}_{corr}(\phi_L\otimes\rho_G^{\beta})+\mathcal{R}_{diss}(\phi_L\otimes\rho_G^{\beta}),
\end{align}
where the correction $\mathcal{R}$ and dissipative $\mathcal{R}$ read as 
\begin{align}    
    \mathcal{R}_{corr}(\phi_L\otimes \rho_G^{\beta})&=-\phi_L\otimes i[Q_B,\rho_G^{{\beta}}]-\bar{O}\phi_L\otimes i(Q_{AB}+Q_{AB}^{\dagger})\rho_G^{\beta}\nonumber\\
    &\quad+\phi_L \bar{O}\otimes i\rho_G^{\beta}(Q_{AB}+Q_{AB}^{\dagger}),   
    \\\mathcal{R}_{diss}(\phi_L\otimes\rho_G^{\beta})&=\bar{O}\phi_L \bar{O}\otimes B\rho_G^{\beta}B^{\dagger}-\frac{1}{2}\phi_L\otimes\{B^{\dagger}B, \rho_G^{\beta}\}\nonumber\\
    &\quad+\phi_L\bar{O}\otimes A\rho_G^{\beta}B^{\dagger}+\bar{O}\phi_L\otimes B\rho_G^{\beta}A^{\dagger}-\frac{1}{2}\{\bar{O}\otimes(A^\dagger B+B^{\dagger}A), \phi_L\otimes\rho_G^{\beta}\},
\end{align}
and the 'correction' Hamiltonians read as
\begin{align}
   Q_B&=\frac{i}{2}\sum_{i,j}\tanh(\frac{\beta(e_i-e_j)}{4})\langle e_i|_G B^\dag B|e_j \rangle _G\ket{e_i}_G\bra{e_j}_G,\\ 
    Q_{AB}&=\frac{i}{2}\sum_{i,j}\tanh(\frac{\beta(e_i-e_j)}{4})\langle e_i|_G A^\dag B|e_j \rangle_G \ket{e_i}_G\bra{e_j}_G.
\end{align}
Here, $\ket{e}_G$ is the eigenstate of the gauge Hamiltonian.
Applying the triangle inequality and the multiplicativity of the trace norm under tensor products, $\mathcal{R}_{corr}(\phi_L\otimes \rho_G^{\beta})$ and $\mathcal{R}_{diss}(\phi_L\otimes \rho_G^{\beta})$ are bounded by 
\begin{align}    
    \left\|\mathcal{R}_{corr}(\phi_L\otimes \rho_G^{\beta})\right\|_1&\leq\left\|\phi_L\right\|_1 \left\|-i[Q_B,\rho_G^{\beta}]\right\|_1+\left\|\bar{O}\phi_L\right\|_1 \left\|-i(Q_{AB}+Q_{AB}^{\dagger})\rho_G^{\beta}\right\|_1\nonumber\\
    &+\left\|\phi_L \bar{O}\right\|_1 \left\|i\rho_G^{\beta}(Q_{AB}+Q_{AB}^{\dagger})\right\|_1, \label{eq: R_corr original expression}   
    \\\left\|\mathcal{R}_{diss}(\phi_L\otimes\rho_G^{\beta})\right\|_1&\leq\left\|\bar{O}\phi_L \bar{O}\right\|_1 \left\|B\rho_G^{\beta}B^{\dagger}\right\|_1+\left\|-\frac{1}{2}\phi_L\right\|_1\left\|\{B^{\dagger}B, \rho_G^{\beta}\}\right\|_1\nonumber\\
    &\quad+\left\|\phi_L\bar{O}\right\|_1\left\| A\rho_G^{\beta}B^{\dagger}\right\|_1+\left\|\bar{O}\phi_L\right\|_1\left\| B\rho_G^{\beta}A^{\dagger}\right\|_1+\left\|-\frac{1}{2}\bar{O}\phi_L\right\|_1\left\|(A^\dagger B+B^{\dagger}A)\rho_G^{\beta}\right\|_1\nonumber\\
    &\quad+\left\|-\frac{1}{2}\phi_L\bar{O}\right\|_1\left\|\rho_G^{\beta}(A^\dagger B+B^{\dagger}A)\right\|_1, \label{eq: R_diss original expression}  
\end{align}
Note that the trace norm of the logical subsystem is order one with respect to the inverse temperature $\beta$, meaning we only care about the trace norms of the gauge subsystem.

In the following section, we simplify the inequalities for Eqs.~\eqref{eq: R_corr original expression} and~\eqref{eq: R_diss original expression} to two primary terms, $\|B\rho_{G}^{\beta}\|_1$ and $\|B^{\dagger}A\rho_{G}^{\beta}\|_1$, thereby proving Lemma~\ref{lemma: lemma 1} in Appendices~\ref{Appendix: Bounding Q terms} and~\ref{appendix: boundinf R_diss}, respectively. Subsequently, we establish explicit upper bounds for both $\|B\rho_{G}^{\beta}\|_1$ and $\|B^{\dagger}A\rho_{G}^{\beta}\|_1$ in Appendix~\ref{Appendix: bounding B rho and BArho}.

\subsection{Bounding the correction Hamiltonians}\label{Appendix: Bounding Q terms}
In this section, we wish to prove that $\norm{Q_M \rho_G^\beta}_1 \leq \norm{M \rho_G^\beta}_1$ and $\norm{\rho_G^\beta Q_M}_1 \leq \norm{\rho_G^\beta M}_1$ for $M \in \{ A^\dagger_q B_q, A_p^\dagger B_p\}$ and $Q_B=0$. 
This result leads to $\left\|\mathcal{R}_{corr}(\phi_L\otimes \rho_G^{\beta})\right\|_1$ being bounded by $\norm{\rho_G^\beta M}_1$ and is a cornerstone of Lemma.~\ref{lemma: lemma 1}. 

\subsubsection{Evaluation of $Q_B$}
Consider $B^\dagger B$, we obtain
\begin{align}
    B^\dagger_q B_q&= \frac{1}{(\Delta y)^2}\int_{-\pi}^{\pi}dx\,\Big(\|J^{+}_{B,q}\|^2\ket{x,1}_G\bra{x,1}_G+\|J^{-}_{B,q}\|^2\ket{x,M}_G\bra{x,M}_G\Big), \label{eqn: BB q}\\
    B^\dagger_p B_p&=\frac{1}{(\Delta x)^2}\int_{-1/4}^{1/4} dy \,\Big(\|J^{+}_{B,p}\|^2\ket{1,y}_G\bra{1,y}_G+\|J^{-}_{B,p}\|^2\ket{N/2,y}_G\bra{N/2,y}_G\Big)\label{eqn: BBp}.
\end{align}
Note that the above expressions are diagonal, and the difference between energies vanishes. 
As a result, the corresponding correction Hamiltonian vanishes,
\begin{equation}
    Q_B=0,
\end{equation}
as the hyperbolic tangent vanishes at zero. 

\subsubsection{Evaluation of $Q_{AB}$}
Moving to the case of $Q_{AB}$, we denote $Q_{AB}$ in the $q$ and $p$ direction to be $Q^{q}_{AB}$ and $Q^{p}_{AB}$, respectively.
The products $A_q^{\dagger}B_q$ and $A_p^\dagger B_p$ are evaluated as
\begin{align}
    A^\dagger_q B_q=&\frac{1}{(\Delta y)^2}\int_{-\pi}^{\pi}dx\,\left(J_{B,q}^{-}J_{A,q,1}^{-}e^{\tfrac{ix}{2}}\ket{x,2}_G\bra{x,M}_G+J_{B,q}^{+}J_{A,q,M-1}^{+}e^{-\tfrac{ix}{2}}\ket{x,M-1}_G\bra{x,1}_G\right),\\
    A_p^\dagger B_p=&\frac{1}{(\Delta x)^2}\int_{-1/4}^{1/4} dy \, \left(J_{B,p}^{-}J_{A,p,1}^{-}\ket{2,y}_G\bra{N/2, y}_G+J_{B,p}^{+}J_{A,p,N/2-1}^{+}\ket{N/2-1,y}_G\bra{1,y})_G\right)\nonumber\\
    &+\frac{2\pi \sqrt{\gamma(0)}}{\Delta x}\int_{-1/4}^{1/4} dy \, \left(J_{B,p}^{+}y\ket{N/2,y}_G\bra{1,y}_G-J_{B,p}^{-}y\ket{1,y}_G\bra{N/2,y}_G\right).
\end{align}
First, note that all of these operators take the following form:
\begin{align}
M = \int d \xi \sum_{i=1}^C \left(\sum_{j=1}^{F} k_{\xi, i, j} \ket{b_{i,j}, \xi}\right) \bra{a_i, \xi} \; ,
\label{eq:form_of_operator_pairs}
\end{align}
where $\{\ket{b_{i,j}, \xi}\}$ and $\{\ket{a_i, \xi}\}$ are two internally orthonormal sets of states parametrized by a shared continuous variable $\xi$ (either $x$ or $y$), and $C$ and $F$ are some numbers less than the dimensionality of the Hilbert space (in this case, $C=2$). As also noted in Appendix \ref{Appendix: bounding B rho and BArho}, the trace norm of such operators turns out to be easy to evaluate. Indeed, we can note that $M^\dagger M$ is a diagonal operator,
\begin{align}
M^\dagger M &= \int d \xi  \, d \xi' \sum_{i,i' = 1}^C \sum_{j, j'=1}^F k_{\xi', i', j'}^* \,  k_{\xi, i, j} \ket{a_{i'}, \xi'} \bra{b_{i', j'}, \xi'}\ket{b_{i, j}, \xi} \bra{a_i, \xi}\\
&= \int d \xi  \, d \xi' \sum_{i,i' = 1}^C \sum_{j, j'=1}^F k_{\xi', i', j'}^* \, k_{\xi, i, j} \, \delta_{i,i'} \delta_{j,j'} \delta(\xi - \xi') \, \ket{a_{i'}, \xi'} \bra{a_i, \xi}\\
&= \int d \xi  \sum_{i = 1}^C \sum_{j=1}^F |k_{\xi, i, j}|^2 \, \ket{a_i, \xi} \bra{a_i, \xi} \; ,
\end{align}
meaning the trace norm is easily read off as 
\begin{align}
\norm{M}_1 = \text{Tr}(\sqrt{M^\dagger M}) = \int d \xi  \sum_{i = 1}^C\left( \sqrt{\sum_{j = 1}^F |k_{\xi, i,j}|^2} \right) .
\end{align}\\
Secondly, note that, since we work in the energy eigenbasis  (i.e., the two sets $\{\ket{\xi, b_i}\}$ and $\{\ket{\xi, a_i}\}$ both consist of elements of the eigenbasis of the Hamiltonian), the form of our operators is actually preserved under multiplication with the thermal state. Indeed, extending the $a$-set to a full basis for the Hilbert space, we can write the thermal state formally as
\begin{align}
\rho_G^{\beta} = \int d \xi \, \sum_{i} \, \frac{e^{- \beta E(a_i, \xi)}}{\mathcal{Z}} \ket{a_i, \xi} \bra{a_i, \xi} \; ,
\end{align}
meaning the product takes the form
\begin{align}
M \rho_G^{\beta} &= \left(\int d \xi \sum_{i=1}^C \left(\sum_{j=1}^{F} k_{\xi, i, j} \ket{b_{i,j}, \xi}\right) \bra{a_i, \xi}\right) \left( \int d \xi' \, \sum_{s} \, \frac{e^{-\beta  E(a_s, \xi')}}{\mathcal{Z}} \ket{a_s, \xi'} \bra{a_s, \xi'}\right)\\
&=  \int d \xi \sum_{i=1}^C \sum_{j=1}^{F} \frac{e^{- \beta E(a_i, \xi)}}{\mathcal{Z}} k_{\xi, i, j}  \ket{b_{i,j}, \xi} \bra{a_i, \xi} \; .
\end{align}
Comparing to Eq. \eqref{eq:form_of_operator_pairs}, we see that this is of the same form except for a simple rescaling $k_{\xi, i, j} \rightarrow \frac{e^{- \beta E(a_i, \xi)} }{\mathcal{Z}} k_{\xi, i,j}$. A similar rescaling applies under multiplication by $\rho_G^\beta$ from the left, with the corresponding rescaling taking the form by $k_{\xi, i,j} \rightarrow \frac{e^{- \beta E(b_{i,j}, \xi)} }{\mathcal{Z}} k_{\xi, i, j}$.\\

Finally, explicit calculation similarly shows that the correction Hamiltonian corresponding to $M$, $Q_M$, takes the same form, with the only difference being a rescaling
\begin{align}
k_{\xi, i, j} \overset{M \text{ to } Q_M}{\longrightarrow}  \frac{i}{2} \tanh\left( \frac{\beta ( E(b_{i,j}, \xi) - E(a_i, \xi))}{4} \right) k_{\xi, i, j} \; .
\end{align}
Note that  since
\begin{align*}
 \left| \frac{i}{2} \tanh\left( \frac{\beta ( E(b_{i,j}, \xi) - E(a_i, \xi))}{4} \right) \right| \leq 1 \; ,
\end{align*}
this implies that
\begin{align}
\norm{Q_M}_1 &= \int d \xi  \sum_{i = 1}^C  \sqrt{\sum_{j = 1}^F\left| \frac{i}{2} \tanh\left( \frac{\beta ( E(b_{i,j}, \xi) - E(a_i, \xi))}{4} \right) k_{\xi, i,j}\right|^2} \\
&\leq \int d \xi  \sum_{i = 1}^C\left( \sqrt{\sum_{j = 1}^F |k_{\xi, i,j}|^2} \right) = \norm{M}_1 \;.
\end{align}
Combining these two rescaling results, one readily sees that
\begin{align}
\norm{Q_M \rho_G^{\beta}}_1 = & \, \int d \xi  \sum_{i = 1}^C \sqrt{\sum_{j=1}^{F}\left|\frac{i}{2} \tanh\left( \frac{\beta ( E(b_{i,j}, \xi) - E(a_i, \xi))}{4} \right) \frac{e^{- \beta E(a_i, \xi)} }{\mathcal{Z}} k_{\xi, i, j} \right|^2} \\
& \leq \int d \xi  \sum_{i = 1}^C \sqrt{\sum_{i = 1}^F \left| \frac{e^{-\beta E(a_i, \xi)} }{\mathcal{Z}} k_{\xi, i, j} \right|^2} = \norm{M \rho_G^{\beta}}_1 \; ,
\end{align}
as desired. Similarly, the properties straightforwardly combine to yield $\norm{\rho_G^{\beta} Q_M}_1 \leq \norm{\rho_G^{\beta} M}_1$.

\subsection{Bounding $\left\|\mathcal{R}_{diss}(\phi_L\otimes\rho_G^{\beta})\right\|_1$ and proof of Lemma~\ref{lemma: lemma 1}}\label{appendix: boundinf R_diss}
In the case of $\left\|\mathcal{R}_{\text{diss}}(\phi_L \otimes \rho_G^{\beta})\right\|_1$, most terms can be expressed and bounded as follows:
\begin{align}
    \left\|T\rho_G^{\beta}B^\dagger\right\|_1,\left\|B\rho_G^{\beta}T^\dagger\right\|_1, \left\|\rho_G^{\beta}B^\dagger T\right\|_1, \left\|T^{\dagger}B\rho_G^{\beta}\right\|_1&\leq\left\|T\right\|_{\infty}\left\|B\rho_G^{\beta}\right\|_1,
\end{align}
These bounds are derived using Hölder's inequality and the fact that the trace norm is invariant under the adjoint operation. Here, $T$ represents any of the operators $A, B, A^\dagger$, or $B^\dagger$.
The remaining terms in $\mathcal{R}_{diss}$ can be written as $\|B^\dagger A\rho_G^\beta\|$, where applying Hölder's inequality does not give us an advantage. 
Combining with the result in Appendix~\ref{Appendix: Bounding Q terms}, we have 
\begin{equation}
        \|\mathcal{R}(\phi_L\otimes\rho_G^\beta)\|_1\leq C \|B \rho_G^\beta\|_1+D\|B^\dagger A\rho_G^\beta\|_1,
\end{equation}
where $C$ and $D$ are constants. In the ideal GKP calculation, the constant $C$ depend on quantities such as $\|A\|_{\infty}$ and $\|B\|_{\infty}$, which can scale as $1/\Delta x$ or $1/\Delta y$. This apparent divergence is a consequence of the infinite-energy nature of the ideal GKP Hamiltonian, rather than a physical feature of the finite-energy problem. In a realistic finite-energy setting, as shown in Appendix.~\ref{Appendix: finite energy simulation}, the GKP peaks acquire a finite width and these singular factors are regularized 
Thus, the physically relevant contribution is the Arrhenius suppression contained in the boundary terms $\|B\rho_G^\beta\|_1$ and $\|B^\dagger A\rho_G^\beta\|_1$.

\subsection{Bounding $\|B\rho_G^{\beta}\|_1$ and $\|B^\dagger A\rho_G^\beta\|_1$}\label{Appendix: bounding B rho and BArho}
 We evaluate $\|B\rho_G^{\beta}\|_1$ and $\|B^\dagger A\rho_G^\beta\|_1$ for both $q$ and $p$ by straight-forward calculation from the definition of Schatten 1 norm, i.e., $\|F\|_1=\Tr(\sqrt{F^\dagger F})$ for an abritary matrix $F$.
We first note that $\|B_p^\dagger A_p \rho_G^{\beta}\|$ is bounded by 
\begin{align}
    \|B_p^\dagger A_p \rho_G^{\beta}\|_1=\|B_p^\dagger (\tilde{A}_p+D_y) \rho_G^{\beta}\|_1\leq\|B_p^\dagger \tilde{A}_p\rho_G^{\beta}\|_1+\|B_p^\dagger D_y \rho_G^{\beta}\|_1.
\end{align}
Considering $\|B^\dagger A\rho_G^\beta\|_1$ for both $q$ and $p$, we then get the following expressions by direct matrix multipication,
\begin{align}
     A_q^\dagger B_qB_q^\dagger A_q&=\frac{1}{(\Delta y)^4}\int_{-\pi}^{\pi}dx\,\Big((J_{B,q}^{-}J_{A,q,m}^{-})^2\ket{x,2}_G\bra{x,2}_G +(J_{B,q}^{+}J_{A,q,m}^{+})^2\ket{x,M-1}_G\bra{x, M-1}_G\Big),\label{eqn: ABBA q}\\
     \tilde{A}_p^\dagger B_pB_p^\dagger \tilde{A}_p&=\frac{1}{(\Delta x)^4}\int_{-1/4}^{1/4}dy\,\Big((J_{B,p}^{-}J_{A,p,m}^{-})^2\ket{2, y}_G\bra{2, y}_G+(J_{B,p}^{+}J_{A,p,m}^{+})^2\ket{N/2-1,y}_G\bra{N/2-1,y}_G\Big)\label{eqn: ABBA p},\\
     D_y^\dagger B_pB_p^\dagger D_y&=\frac{4\pi^2\gamma(0)}{(\Delta x)^2}\int_{-1/4}^{1/4} dy\Big(\|J_{B,p}^{+}\|^2y^2\ket{N/2,y}_G\bra{N/2,y}_G+\|J_{B,p}^{-}\|^2y^2\ket{1,y}_G\bra{1,y}_G\Big) \label{eqn: DBBD}.
\end{align}

Since Eqn.~\eqref{eqn: BB q},~\eqref{eqn: BBp}, ~\eqref{eqn: ABBA q}, ~\eqref{eqn: ABBA p} and \eqref{eqn: DBBD} are diagonal, their singular values are easy to read off.
At small $\Delta x,\Delta y$, the jump rates collapse to $\pm i\pi\sqrt{\gamma(0)}$ and $\cos{(x+2\Delta x)}\approx \cos{(x)}$, giving the formal expressions
\begin{align}
    \sqrt{\rho_G^{\beta}A_q^\dagger B_qB_q^\dagger A_q\rho_G^{\beta}} &\sim\frac{1}{(\Delta y)^2}\int_{-\pi}^{\pi}dx\,\frac{e^{\beta E_J(-1+\cos{(x)})}}{\mathcal{Z}}\Big(\ket{x,2}_G\bra{x,2}_G+\ket{x,M-1}_G\bra{x,M-1}_G\Big) \label{eq:sqrt_number_1}\\
    \sqrt{\rho_G^{\beta}B^\dagger_q B_q\rho_G^{\beta}}&\sim\frac{1}{\Delta y}\int_{-\pi}^{\pi}dx\,\frac{e^{\beta E_J(-1+\cos{(x)})}}{\mathcal{Z}}\Big(\ket{x,1}_G\bra{x,1}_G+\ket{x,M}_G\bra{x,M}_G\Big) \label{eq:sqrt_number_2}\\
    \sqrt{\rho_G^{\beta}\tilde{A}_p^\dagger B_pB_p^\dagger \tilde{A}_p\rho_G^{\beta}}&\sim \frac{1}{(\Delta x)^2}\int_{-1/4}^{1/4} dy \,\frac{e^{\beta E_J(-1+\cos{(4\pi y)})}}{\mathcal{Z}}\Big(\ket{2,y}_G\bra{2,y}_G+\ket{N/2-1,y}_G\bra{N/2-1,y}_G\Big) \label{eq:sqrt_number_3}\\
    \sqrt{\rho_G^{\beta} B^\dagger_p B_p\rho_G^{\beta}}&\sim \frac{1}{\Delta x}\int_{-1/4}^{1/4} dy \,\frac{e^{\beta E_J(-1+\cos{(4\pi y)})}}{\mathcal{Z}}\Big(\ket{1,y}_G\bra{1,y}_G+\ket{N/2,y}_G\bra{N/2,y}_G\Big) \label{eq:sqrt_number_4}\\
    \sqrt{\rho_G^{\beta}D_y^\dagger B_pB_p^\dagger D_y\rho_G^{\beta}}&\sim\frac{1}{\Delta x}\int_{-1/4}^{1/4}dy\,\frac{\abs{y}e^{\beta E_J(-1+\cos{(4\pi y)})}}{\mathcal{Z}}\Big(\ket{N/2,y}_G\bra{N/2,y}_G+\ket{1,y}_G\bra{1,y}_G\Big).
\end{align}
Performing a trace and expanding the partition functions in Eqn. \eqref{eq:sqrt_number_1} and \eqref{eq:sqrt_number_2}, we get integrals of the following form, 
\begin{align} 
    \frac{ e^{-\beta E_J} \int_{-\pi}^{\pi} dx\, e^{\beta E_J \cos x}}
    { \int_{-1/4}^{1/4} dy\, e^{\beta E_J \cos(4\pi y)}
    \int_{-\pi}^{\pi} dx\, e^{\beta E_J \cos x}} 
    &=
    e^{-\beta E_J}
    \frac{1}{\int_{-1/4}^{1/4} dy\, e^{\beta E_J \cos(4\pi y)}} \\
    &=
    e^{-\beta E_J}
    \frac{2}{I_0(\beta E_J)} \,,
\end{align}
 where $I_0$ is the modified Bessel function of the first kind. Similarly, Eqn. \eqref{eq:sqrt_number_3} and \eqref{eq:sqrt_number_4} give integrals of the form 
\begin{align}
    \frac{ e^{-\beta E_J}\int_{-1/4}^{1/4} dy\, e^{\beta E_J \cos(4\pi y)}}
    { \int_{-1/4}^{1/4} dy\, e^{\beta E_J \cos(4\pi y)}
    \int_{-\pi}^{\pi} dx\, e^{\beta E_J \cos x}} &=
    e^{-\beta E_J}
    \frac{1}{ \int_{-\pi}^{\pi} dx\, e^{\beta E_J \cos x}} \\
    &=
    e^{-\beta E_J}
    \frac{1}{2\pi I_0(\beta E_J)} \,,
\end{align}
It is known that $I_0(\beta E_J)$ asymptotically approaches $\frac{e^{\beta E_J}}{\sqrt{\beta E_J}}$ as $\beta E_J$ approaches infinity. The calculation therefore gives
\begin{align}
    \|B_q^{\dagger}A_q\rho_G^{\beta}\|_1 &\propto \frac{1}{(\Delta y)^2}\sqrt{\beta E_J}e^{-2\beta E_J},\\
    \|B_p^{\dagger}\tilde{A}_p\rho_G^{\beta}\|_1 &\propto \frac{1}{(\Delta x)^2}\sqrt{\beta E_J}e^{-2\beta E_J},\\
    \|B_q\rho_{G}^{\beta}\|_1 &\propto \frac{1}{\Delta y}\sqrt{\beta E_J}e^{-2\beta E_J},\\
    \|B_p\rho_{G}^{\beta}\|_1 &\propto \frac{1}{\Delta x}\sqrt{\beta E_J}e^{-2\beta E_J}.
\end{align}
For the case of $\|B_p^{\dagger}D_y\rho_{G}^{\beta}\|_1$, we obtain
\begin{align}
    \|B_p^{\dagger}D_y\rho_{G}^{\beta}\|_1&=\Tr\left[\sqrt{\rho_G^{\beta}D_y^\dagger B_pB_p^\dagger D_y\rho_G^{\beta}}\right]\\
    &\propto \frac{1}{\Delta x}\frac{ e^{-\beta E_J}\int_{-1/4}^{1/4} dy\, \abs{y} e^{\beta E_J \cos(4\pi y)}}
    { \int_{-1/4}^{1/4} dy\, e^{\beta E_J \cos(4\pi y)}
    \int_{-\pi}^{\pi} dx\, e^{\beta E_J \cos x}}
\end{align}
Using $\cos(\theta) \leq 1 - \frac{2 \theta^2}{\pi^2}$, the numerator integral is written as
\begin{align}
    e^{-\beta E_J}\int_{-1/4}^{1/4} dy\, \abs{y} e^{\beta E_J \cos(4\pi y)}&\leq2\int_{0}^{1/4} dy\, y e^{-32\beta E_Jy^2}\\
    &\leq\int_{0}^{\infty} d(y^2)\, e^{-32\beta E_J y^2}\\
    &=\frac{1}{32 \beta E_J}.
\end{align}
Since $I_0(\beta E_J)$ asymptotically approaches $\frac{e^{\beta E_J}}{\sqrt{\beta E_J}}$ as $\beta E_J$ approaches infinity, we get
\begin{align}
    \|B_p^{\dagger}D_y\rho_{G}^{\beta}\|_1&=\Tr\left[\sqrt{\rho_G^{\beta}D_y^\dagger B_pB_p^\dagger D_y\rho_G^{\beta}}\right]\\
    &\propto \frac{1}{\Delta x}\frac{\beta E_J e^{-2\beta E_J}}
    {\beta E_J}\\
    &=\frac{1}{\Delta x}e^{-2\beta E_J}
\end{align}
As a result, we have
\begin{equation}
    \|B_p^{\dagger}A_p\rho_G^{\beta}\|_1\leq\|B_p^{\dagger}\tilde{A}_p\rho_G^{\beta}\|_1+\|B_p^{\dagger}D_y\rho_G^{\beta}\|_1\sim\left(\frac{\sqrt{\beta E_J}}{(\Delta x)^2}+\frac{1}{\Delta x}\right) e^{-2\beta E_J}
\end{equation}

\section{Finite energy simulation}\label{Appendix: finite energy simulation}

In this appendix, we numerically verify the modified Arrhenius form of the boundary norms for the finite-energy GKP Hamiltonian. Our goals are twofold. First, we show that the $\frac{1}{\Delta}$ divergence that appears in Appendix~\ref{Appendix: proof of lemma 1} is an artifact of working with the ideal GKP Hamiltonian, whose eigenstates have infinite energy; in the finite-energy setting, the boundary norm instead converges. Second, we demonstrate the Arrhenius-law temperature dependence of the boundary norm. Although our analytical treatment focuses on the ideal GKP limit because it has a clean subsystem-code structure and permits a transparent derivation of the Arrhenius form, the finite-energy model provides the physically relevant setting. 

In our simulations, we consider the Hamiltonian
\begin{equation}
H_{E_h,\mathrm{GKP}}
= E_h\left(\left(\frac{1}{D}\right)^2q^2+mp^2\right)-E_J\bigl(\cos(q)+\cos(4\pi p)\bigr) 
\end{equation}
where $E_h$ is the strength of a harmonic trap, {$D$ and $m$ is some parameters related to the scales of the confinement of the $q$ and $p$ quadratures, respectively}. The resulting approximate GKP states are no longer mutually orthogonal Dirac combs, but rather can be understood as a train of Gaussian peaks whose width is controlled by $\kappa^{1/4}=(E_h/E_J)^{1/4}$~\cite{nathan2025selfcorrectinggkpqubitgates}. Thus, smaller $\kappa$ corresponds to a closer approximation to the ideal GKP limit, while finite energy is retained as long as $\kappa$ remains non-zero~\cite{PhysRevLett.125.040501}. Numerically, we evaluate $\|(\mathbb{I}\otimes O)\rho^{\beta}\|_1$, where $O$ is a boundary operator acting on the gauge subsystem and $\rho^{\beta}$ is the thermal state of the full Hamiltonian. 
To extract $\mathbb{I}\otimes O$, we decompose the jump operator into ideal logical Pauli components using projectors, such that
\begin{equation}
    L_j=\mathbb{I}\otimes A_j+\sum_{\sigma\in\{\bar{X}, \bar{Y}, \bar{Z}\}} \sigma\otimes B_{j, \sigma}.
\end{equation} 
Note that this decommposition implicitly depends on the Hamiltonian, and thus on $\kappa$, through the thermal rewrighting of Eq. \eqref{eq: expression for original jump opertor}. However, when $\kappa$ is sufficiently small, the decomposition approximates the subsystem-code structure of the ideal GKP limit~\cite{PhysRevLett.125.040501}. For example, the position jump operator takes the approximate form $L_q\simeq\mathbb{I}\otimes A_q+\bar{Z}\otimes B_{q,\bar{Z}}$, with corrections that vanish as the finite-energy code approaches the ideal limit.
We further assume that the thermal state is approximately supported on this subsystem structure,
\begin{equation}
\rho^{\beta}\approx \mathbb{I}\otimes \rho_G^{\beta},
\end{equation}
so that the computed quantity well approximates $\|O\rho_G^\beta\|_1$. These approximations deteriorate when $\kappa$ is too large, i.e., far from the ideal GKP limit, and therefore our simulations are performed at small $\kappa$ and large $E_J$. The latter choice corresponds to working in the low-temperature limit, where the subsystem structure is expected to be more robust and numerical finite-size effects from real-space domain boundaries less pronounced.\\
In this finite-energy setting, the boundary norm is modified by vacuum-fluctuation effects, which account for leakage of the imperfect approximate GKP states, as discussed in~\cite{nathan2025selfcorrectinggkpqubitgates}. We therefore expect boundary operators such as $B_qA_q$ or $B_q$ to obey a finite-energy-corrected Arrhenius bound of the form
\begin{equation}\label{eq: finite-energy-corrected Arrhenius form}
    \|O\rho_G^\beta\|_1,\ \|\rho_G^\beta O\|_1 \lesssim C_0 \sqrt{\beta E_J}\exp\left[\frac{-E_J} {C_T k_B T + C_1\sqrt{\kappa}E_J} \right],
\end{equation}
for suitable constants $C_0$, $C_T$, and $C_1$.

\subsection{Convergence of boundary norms}
We first test the convergence of the boundary norm in the finite-energy model. Throughout this subsection, we display the representative quantity $\|B_q\rho_G^\beta\|_1$; we have verified that the other boundary norms used in the analysis exhibit the same qualitative behaviour. 
The parameters are fixed to
\begin{equation}
E_J=50,\qquad \beta=1,
\end{equation}
and we scan over $\kappa=5\times10^{-5}$ to $\kappa=2\times10^{-4}$. For each value of $\kappa$, the boundary norm is evaluated as a function of the inverse resolution parameter $\frac{1}{\Delta}$.
\begin{figure}
    \centering
    \includegraphics[width=0.7\linewidth]{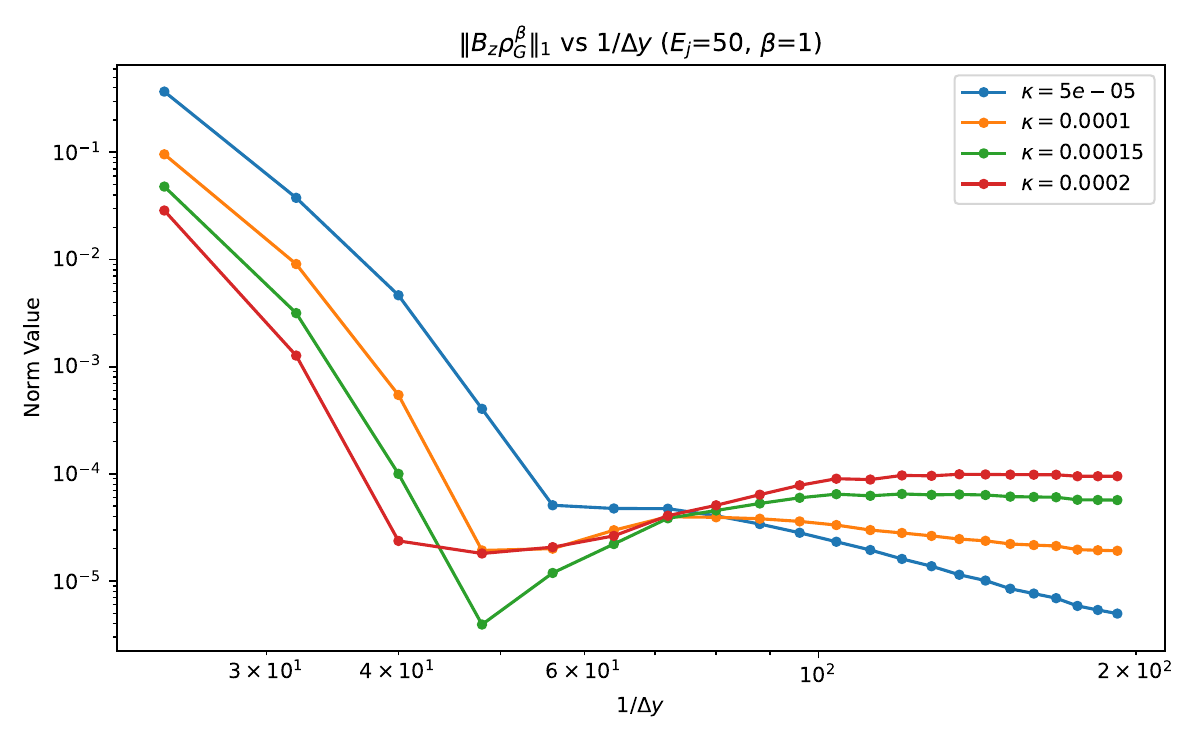}
    \caption{Convergence of the representative boundary norm $\|B_q\rho^\beta_G\|_1$ for $E_J=50$ $\kappa=10^{-4}$, and $\beta=1$ at different values of $\kappa=5\times10^{-5},10^{-4},1.5\times10^{-4}$, and $2\times10^{-4}$. The finite-energy boundary norm saturates as $\Delta$ is decreased, in contrast to the $1/\Delta$ divergence obtained in the ideal GKP limit.}
    \label{fig:boundary-convergence}
\end{figure}

The log-log plot in Fig.~\ref{fig:boundary-convergence} shows that $\|B_q\rho_G^\beta\|_1$ converges as $\Delta$ is decreased. In particular, the finite-energy result does not follow the $\frac{1}{\Delta}$ divergence found in the analytical calculation, confirming that this divergence is removed once the GKP states are regularized by finite-energy effects. We also observe that the convergence improves as $\kappa$ is decreased, consistent with the fact that smaller $\kappa$ places the system closer to the ideal GKP limit while still retaining finite energy.

\subsection{Verification of the finite-energy-corrected Arrhenius form}

\begin{figure}
    \centering
    \includegraphics[width=0.7\linewidth]{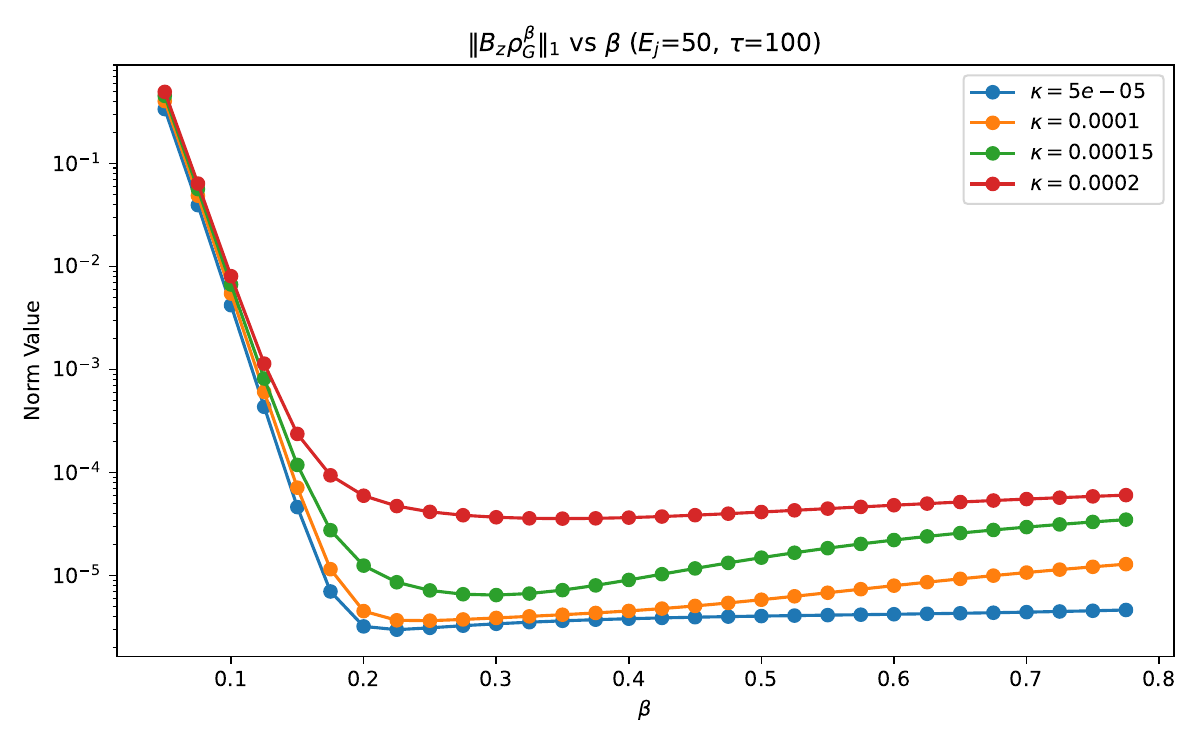}
    \caption{Temperature dependence of the representative boundary norm $\|B_q\rho_G^\beta\|_1$ at fixed system size and $E_J=50$ at different values of $\kappa=5\times10^{-5},10^{-4},1.5\times10^{-4}$, and $2\times10^{-4}$. The semi-log plot shows an approximately Arrhenius decay at intermediate $\beta$, followed by a low-temperature saturation consistent with the finite-energy vacuum-fluctuation correction.}
    \label{fig:boundary-temperature}
\end{figure}

We next study the temperature dependence of the boundary norm. For fixed system size and $E_J=50$, we scan over inverse temperature $\beta$ at different values of $\kappa=5\times10^{-5},10^{-4},1.5\times10^{-4}$, and $2\times10^{-4}$ and compute the representative quantity $\|B_q\rho^\beta\|_1$. The result is shown in Fig.~\ref{fig:boundary-temperature} on a semi-log scale.

At sufficiently high temperature, Eq.~\eqref{eq: finite-energy-corrected Arrhenius form} reduces to the usual Arrhenius behavior
\begin{equation}
    \|O\rho_G^\beta\|_1\sim C_0\exp\left[-\frac{1}{C_T}\beta E_J\right],
\end{equation}
so that $\log\|O\rho^\beta_G\|_1$ is approximately linear in $\beta$. This is the behaviour observed in the intermediate-temperature regime of Fig.~\ref{fig:boundary-temperature}.

At low temperature, however, the finite-energy correction becomes important. Taking the zero-temperature limit $T\to0$, or equivalently $\beta\to\infty$, gives
\begin{equation}
\|O\rho^{\beta}\|_1,\ \|\rho^{\beta}O\|_1
\lesssim
C_0 \sqrt{\beta E_J} \exp \left(\frac{-1}{C_1\sqrt{\kappa}}\right).
\end{equation}
Thus the boundary norm is not expected to decreasse indefinitely with decreasing temperature. Instead, it first approaches a residual finite-energy floor set by the finite width of the approximate GKP peaks, and then exhibits a very slow increase due to the weak $\sqrt{\beta}$ prefactor. The flattening, followed by a slight upward trend at large $\beta$ in Fig.~\ref{fig:boundary-temperature}, is therefore consistent with the vacuum-fluctuation correction in the finite-energy model, rather than a failure of the Arrhenius picture.

\section{Corrections from physical dissipation models}\label{Appendix: Corrections from physical dissipation models}
The robustness discussed in the main text is {\it perturbatively stable}, and applies for finite corrections to the dissipator above.
We now comment on the corrections to the above result from physical dissipation models. 
A physically derived dissipation model can also be established for the model, by coupling the system degree of freedom $S^a $ to a bath $\mathcal B_a$ with power spectral density $\gamma(\omega)$.
This shares the same jump operator as the above, but has a different Hamiltonian. 
\begin{equation}\label{eq: CKG lindadian 2}
    \begin{split}
        \mathcal{L}(\rho) = -i \left[ H+\Lambda, \rho \right]+ \sum_a L_a \rho L_a^{\dagger} - \frac{1}{2} \left( L_a^{\dagger} L_a \rho + \rho L_a^{\dagger} L_a \right),
    \end{split}
\end{equation}
with $\Lambda $ a local Hamiltonian coupling neighbouring sites, of order $\Gamma$: $
    \Lambda = \sum_{jkl}|e_j\rangle\langle e_k| S^a _{jl}S^a_{lj} f(E_{lj},E_{kl})
$
and $f(\epsilon_1,\epsilon_2)=-2\pi  \mathcal P \int d\omega \omega^{-1}g(\omega-\epsilon_1,\omega+\epsilon_2)$, and $g(\omega) = \sqrt{\gamma(\omega)/2\pi}$ . 
The analsyis can also be applied to this Lindbladian: the only difference in the analysis is that  the Gauge steady state $\rho_{\rm G}$ will be modified to a dressed Gibbs state to account for physical finite-coupling corrections~\cite{nathan2020quantifying,thingna_generalized_2012,nathan2020responsecommentuniversallindblad}  to the steady state. Importantly, the support of the dressed state at the edge of the torus will resemble a Gibbs state and remain of order $e^{-2\beta E_J}$, while the support of $B$ will still just be at the edge of the torus.
Thus the exponentially scaling lifetime persists for physically motivated dissipators.

We can extend this argument to even include non-Markovian corrections, to make our arguments exact for a given microscopic model. 
By including the bath in the  gauge  subsystem, the Liouvillain of the combined system reads 
\begin{equation}
\begin{aligned}
    \mathcal L_{\rm exact} = -i\Big[1\otimes\big( H_S\otimes 1+ \sum_aS^a_1\otimes \mathcal B\big),\cdot \Big]   -i\Big[\sum_a \bar O_a \otimes  S_a ^2 \otimes \mathcal B_a,\cdot \Big].
    \end{aligned}
\end{equation}
Here  $S_1^a $ and $S_2^a $ denote the components of the quadrature  $a$ that act within the torus, and on the edge, respectively.
We let $\rho_0 $ be the steady state of the first two terms of the Liouvillian: this will resemble a Gibbs state and can be obtained systematically through canonical perturbation theory or expansions of the dissipator~\cite{thingna_generalized_2012,agerskov2026markovianquantummasterequations}. 
We then have 
\begin{equation}
\begin{aligned}
    \norm{\mathcal L_{\rm exact}[\phi_L\otimes\rho_{0}]}_{\rm tr}  = \norm{\sum_a[\bar O_a \otimes S_2^a \otimes \mathcal B_a,\rho_0]}_{\rm tr}  \leq  2\sum_a \norm{(\bar O_a \otimes S_2^a \otimes \mathcal B_a)(\phi_L\otimes \rho_0)}_{\rm tr} 
    \end{aligned}
\end{equation}
Using that $\norm{X}_{\rm tr}=\Tr[U |X|  ]$ for some unitary $U$, along with the Cauchy-Schwarz inequality, $\Tr[\rho AB]\leq \sqrt{\Tr[\rho A^2 ]\Tr[\rho B^2]}$~\cite{agerskov2026markovianquantummasterequations}, we find 
\begin{equation}
    \norm{\mathcal L_{\rm exact}[\phi_L\otimes\rho_{0}]}_{\rm tr} \leq    2\sum_a \sqrt{ \Tr[ (S^a _2 ) ^2 \rho_0]}\Gamma _a 
\end{equation}
With $\Gamma_a  = \sqrt{\Tr[\rho_0\mathcal B_a^2]}$ a finite rate characterizing the magnitude of bath fluctuations. Now, since $S_j'$ has only support near the edges of the zone, and $\rho_0$ resembles a Gibbs state, we find  
\begin{equation}
    \sqrt{ \Tr[( S_2^a)^2 \rho_0]}\sim e^{-2 \beta E_J}
\end{equation}
Thus,
    \begin{equation}
        \norm{\mathcal L_{\rm exact}[\phi_L\otimes\rho_{0}]}_{\rm tr} \lesssim \Gamma  e^{-2 \beta E_J}.
    \end{equation}
    With $\Gamma = \max(\Gamma_p,\Gamma_q)$.
    Thus, the qubit is exponentially long-lived even when accounting for non-Markovian effects in a microscopic dissipation model.

Importantly, the same line of arguments can be extended to establish the exponential robustness in the presence of finite perturbations to the Hamiltonian itself. The only condition to be satisfied is that the Hamiltonian consists of terms that are either local (i.e., finite-order polynomial of quadratures), or are periodic functions of $p$ and $q$ (specifically, explicit functions of $\bar x$ and $\bar p$, or equivalently,  commute with the GKP stabilizers), such that there is an energy penalty associated with the boundaries of the Zak torus.

\end{document}